\documentclass[lettersize,journal]{IEEEtran}
\usepackage{amsfonts,dsfont,amssymb,amsbsy,amsmath,paralist,theorem,bm,ifthen,color}
\usepackage[pdfstartview=FitH,bookmarksnumbered,unicode,bookmarksopen=true]{hyperref}
\usepackage{graphicx}
\usepackage{epstopdf}
\usepackage{relsize}
\usepackage{mathrsfs}
\usepackage{tcolorbox}
\usepackage[caption=false,font=footnotesize]{subfig}
\usepackage{bm}
\usepackage{stfloats}
\usepackage{url}
\usepackage[noadjust]{cite}
\usepackage{cases,booktabs}
\usepackage{graphicx,algorithmic,algorithm,relsize}
\usepackage[justification=centering]{caption} 
\usepackage{array,tabularx, multirow, multicol, booktabs, hyperref}
\usepackage{threeparttable}
\usepackage{tablefootnote}

\newtheorem{lem}{Lemma}
\newtheorem{prop}{Proposition}

\newtheorem{rem}{Remark}

\newcommand\Mset  {\ensuremath{{\mathcal{M}}}}

\newcommand\st    {\ensuremath{{\rm s.t.}}}

\graphicspath{{fig/}}

\definecolor{grey}{RGB}{80 80 80}
\definecolor{green}{RGB}{34	195	46}
\definecolor{red}{RGB}{220 0 0}

\usepackage{graphicx} 
\allowdisplaybreaks[4]

\title{Pinching-Antenna System Design under Random LoS and NLoS Channels} 

\author{Yanqing Xu, \IEEEmembership{Member, IEEE,} 
        Yang Lu, \IEEEmembership{Member, IEEE,} 
        Zhiguo Ding, \IEEEmembership{Fellow, IEEE,} \\
        and Tsung-Hui Chang, \IEEEmembership{Fellow, IEEE}
        \thanks{\smaller[1] Part of this work has been submitted to IEEE ICC 2026 \cite{xu2026optimal}.}
        \thanks{\smaller[1] Y. Xu is with the School of Science and Engineering, The Chinese University of Hong Kong (Shenzhen), Longgang, Shenzhen, Guangdong, 518172, P.R. China (email: xuyanqing@cuhk.edu.cn).}
        \thanks{\smaller[1] Y. Lu is with the State Key Laboratory of Advanced Rail Autonomous Operation, and also with the School of Computer Science and Technology, Beijing Jiaotong University, Beijing 100044, China (e-mail:yanglu@bjtu.edu.cn)}
        \thanks{\smaller[1] Z. Ding is with the University of Manchester, Manchester, M1 9BB, UK (email: zhiguo.ding@manchester.ac.uk).}
        \thanks{\smaller[1] T.-H. Chang is with the School of Artificial Intelligence, The Chinese University of Hong Kong, Shenzhen, 518172, China (email: changtsunghui@cuhk.edu.cn).} 
        \thanks{\smaller[1] This work has been submitted to the IEEE for possible publication. Copyright may be transferred without notice, after which this version may no longer be accessible.}\vspace{-0mm}
        }

\date{\today}

\begin{document}

\maketitle

\begin{abstract}
    Pinching antennas, realized through position-adjustable radiating elements along dielectric waveguides, have emerged as a promising flexible-antenna technology thanks to their ability to dynamically reshape large-scale channel conditions. However, most existing studies focus on idealized LoS-dominated environments, overlooking the stochastic nature of realistic wireless propagation. This paper investigates a more practical multiuser pinching-antenna system under a composite probabilistic channel model that captures distance-dependent LoS blockage and NLoS scattering. To account for both efficiency and reliability aspects of communication, two complementary design metrics are considered: an average signal-to-noise ratio (SNR) metric characterizing long-term throughput and fairness, and an outage-constrained metric ensuring a prescribed reliability level. Based on these metrics, we formulate two optimization problems: the first maximizes the max–min average SNR across users, while the second maximizes a guaranteed SNR threshold under per-user outage constraints. Although both problems are inherently nonconvex, we exploit their underlying monotonic structures and develop low-complexity, bisection-based algorithms that achieve globally optimal solutions using only simple scalar evaluations. Extensive simulations validate the effectiveness of the proposed methods and demonstrate that pinching-antenna systems significantly outperform conventional fixed-antenna designs even under random LoS and NLoS channels. 
\end{abstract}

\begin{IEEEkeywords}
     Pinching antenna, line-of-sight blockage, non-line-of-sight channels, average signal-to-noise ratio maximization, outage probability constraint.
\end{IEEEkeywords}

\section{Introduction}
Multi-antenna technologies have played a pivotal role in advancing modern wireless communication systems, enabling high spectral efficiency, wide coverage, and improved link reliability through spatial diversity and beamforming \cite{larsson2014massive,bjornson2023twenty,xu2025distributed}. Despite these advances, conventional antenna arrays are typically deployed at fixed locations, which fundamentally limits their ability to adapt to environmental dynamics such as user mobility and random line-of-sight (LoS) blockage. As a result, static antenna deployments often suffer severe performance degradation in dynamic or obstacle-rich environments, where maintaining robust LoS links becomes increasingly difficult. To overcome these limitations, various flexible and reconfigurable antenna technologies have emerged, including reconfigurable intelligent surfaces (RISs) \cite{liu2021reconfigurable,tang2020wireless}, movable antennas \cite{zhu2023movable,zhang2024channel}, and fluid antennas \cite{wong2020fluid,new2024tutorial}. These paradigms enable the physical or electromagnetic reconfiguration of antenna structures to better adapt to propagation environments. However, they still face fundamental challenges in dynamically reconstructing LoS links and modifying large-scale channel characteristics. Specifically, once RISs are deployed, they can only adjust signal phases rather than change propagation paths, while movable and fluid antennas typically operate over limited spatial ranges on the order of a few wavelengths, restricting their capability to circumvent LoS blockage across extended regions.

Pinching antennas have recently emerged as a promising new direction for flexible-antenna design \cite{suzuki2022pinching,ding2024flexible,liu2025pinching,xu2025generalized}. In this architecture, radiating elements can be activated at arbitrary positions along a dielectric waveguide, allowing the antenna to be effectively pinched and reconfigured along its length. This unique structure provides an additional geometric degree of freedom, i.e., the antenna position, which can be dynamically adjusted according to user locations or environmental conditions. Such adaptability enables the system to actively establish strong LoS paths, mitigate shadowing, and enhance link reliability over large communication areas, offering substantial advantages over traditional antenna system.

Owing to its unique capability of enabling large-scale spatially reconfigurable transmission, pinching-antenna systems have attracted growing research interest. 
The initial work \cite{ding2024flexible} established the electromagnetic and system models of pinching antennas and demonstrated their flexibility in dynamically creating and maintaining strong LoS links with reduced propagation loss. Following these results, the authors of \cite{xu2025rate} studied the downlink data rate maximization problem in a single-waveguide pinching-antenna system and a two-stage algorithm was proposed to simultaneously minimize the large-scale fading and guarantee the constructive signal reception at the user. Then, the studies were extend to the multiuser and multi-waveguide scenarios in \cite{wang2025modeling,bereyhi2025mimo,zhang2025two}, where joint optimization of antenna positions and transmit beamformers was shown to significantly improve spectral and energy efficiencies. 
To further enhance the spectral efficiency, the works \cite{wang2025antenna,xu2025qos,fu2025power} explored non-orthogonal multiple access (NOMA)-assisted pinching-antenna systems, where signals of multiple users are superposed and propagated through the same waveguide to enhance multiplexing gain. Meanwhile, the integration of pinching antennas with emerging paradigms such as integrated sensing and communication (ISAC) was investigated in \cite{ding2025pinching,khalili2025pinching,mao2025multi}, where the flexible antenna positioning was exploited to enhance sensing accuracy and communication coverage simultaneously. 
The works \cite{xu2025pinching,xu2025pinching-los} investigated the impact of in-waveguide attenuation on the performance of pinching antennas in LoS-dominated and random LoS blockage scenarios, respectively.  
Reliability-oriented analyses were conducted in \cite{ding2025los,wang2025pinching_los}, which studied how LoS blockage affects the pinching-antenna system, revealing that dynamic antenna repositioning effectively mitigates LoS blockage and inter-user interference. Recently, using the pinching antenna's capability to reconfigure LoS links, an environment division multiple access (EDMA) technique was proposed in \cite{ding2025EDMA}, showing the advantages of pinching antennas on supporting multi-user communications.
Summarily, while existing research has demonstrated the potential of pinching antennas in improving link quality and spectral efficiency, most of these studies assume deterministic or LoS-dominated channels, thereby overlooking the stochastic nature of realistic wireless propagation environments. In practice, LoS links may be intermittently blocked by obstacles, and NLoS scattering often contributes significantly to overall performance \cite{3gpp2020channel}. Consequently, an accurate performance characterization and optimal design of pinching-antenna systems under random LoS blockage and random NLoS scattering remain open and essential research problems.

\subsection{Contributions}

To bridge the above gap, this paper investigates a realistic channel model that jointly captures random LoS blockage and NLoS fading, and studies the optimal design of multiuser pinching-antenna systems under two complementary performance metrics: an average signal-to-noise ratio (SNR) metric and an outage-constrained metric. Based on these two metrics, we formulate and solve two corresponding optimization problems. The main contributions are summarized as follows:

\begin{itemize}
    \item \textbf{Average-SNR-based design:} We first investigate an average SNR metric that characterizes the long-term throughput and fairness performance of multiuser pinching-antenna systems. Based on this metric, a max-min average SNR optimization problem is formulated to balance user link qualities through antenna placement along the waveguide. The problem is inherently nonconvex due to the nonlinear coupling between user positions and the antenna position. To tackle this challenge, we reformulate the problem in an epigraph form and analyze the structural properties of the average SNR under the composite LoS/NLoS channel model, showing that, for any given SNR threshold, each user's feasibility set in the antenna-position domain is a close interval and that these intervals are nested with respect to the SNR threshold. This reveals that the problem is quasi-convex in the antenna position and enables a globally optimal, low-complexity bisection-based algorithm that updates the SNR threshold while checking feasibility via simple scalar calculations.

    \item \textbf{Outage-constrained design:} We then focus on an outage-constrained metric that targets short-term link reliability. Based on this metric, an outage-constrained SNR maximization problem is formulated with the requirement that each user must satisfy a prescribed outage probability. To solve this problem, we first derive a closed-form expression for the complementary cumulative distribution function (CCDF) of the instantaneous SNR under the proposed random LoS/NLoS model. By exploiting the monotonicity of this CCDF with respect to the user-to-antenna distance, each outage constraint is equivalently transformed into a one-dimensional distance bound obtained from a scalar monotone equation. This structural property leads to a globally optimal bisection-based algorithm, where the outer loop searches over the SNR threshold and the inner loop identifies the feasible antenna-position interval satisfying all outage constraints, while maintaining low computational complexity.

    \item \textbf{Performance insights:} Extensive simulations validate the proposed algorithms and provide quantitative insights into the efficiency-reliability tradeoff in pinching-antenna systems. The results show that pinching antennas substantially outperform conventional fixed-antenna deployments in both long-term and reliability-oriented metrics, even under severe LoS blockage and NLoS fading, and that the performance gain becomes more pronounced in larger coverage regions and under stringent reliability requirements.
\end{itemize}

The rest of the paper is organized as follows. Section~\ref{sec:system_model} presents the system model, the channel model and the two design metrics. Section \ref{sec:avg_snr} investigates the average-SNR-based optimization problem and propose a low-complexity single-loop bisection search-based algorithm. Section \ref{sec:outage} presents the outage-constrained formulation and proposed a double-loop bisection search-based algorithm. Section~\ref{sec:simulation} provides numerical results and performance discussions, and Section~\ref{sec:conclusions} concludes the paper.


\section{System Mode and Channel Model} \label{sec:system_model}

\subsection{System Model}
We consider a downlink system where a single waveguide equipped with one pinching antenna serves $M$ single-antenna users, indexed by $m \in \mathcal{M} \triangleq \{1, \ldots, M\}$.
The users are randomly distributed within a rectangular service region of size $D_x \times D_y$. The position of user $m$ is denoted by $\boldsymbol{\psi}_m = [x_m, y_m, 0]^{\top}$, where $0 \leq x_m \leq D_x$ and $-D_y/2 \leq y_m \leq D_y/2$.
The waveguide is deployed parallel to the $x$-axis at a height $d_v$ above the user plane. The pinching antenna is located at $\widetilde{\boldsymbol{\psi}}^{\mathrm{pin}} = [\tilde{x}, 0, d_v]^{\top}$, where $\tilde{x} \in [0, D_x]$ is a design variable to be optimized.


The orthogonal multiple access (OMA) scheme is adopted, such as time-division multiple access (TDMA) or orthogonal frequency-division multiple access (OFDMA). In OMA, each user is allocated an orthogonal transmission resource (time slot or frequency subband), ensuring interference-free communication among users. In this work, we assume that the pinching-antenna position remains fixed during each transmission frame \cite{ding2025analytical}. This assumption is practically reasonable for both TDMA and OFDMA: in TDMA systems, dynamically repositioning the antenna in every short time slot would introduce excessive mechanical complexity and latency; in OFDMA systems, changing the antenna position within a frame would break the subcarrier orthogonality required for interference-free transmission. Without loss of generality, we focus on the TDMA-based OMA scheme in the remainder of this paper.

\subsection{Channel Model with Random LoS and NLoS Components} 

The wireless channel between the pinching antenna and user $m$ is modeled as a combination of a probabilistic LoS component and an NLoS component, in accordance with modern mmWave channel modeling \cite{3gpp2020channel,poddar2023tutorial}. Specifically, We employ a hybrid model capturing both near-field spherical-wave propagation for the LoS component and a cluster-based statistical fading model for the NLoS component.

\subsubsection{Probabilistic LoS Channel Component}
Let $r_m = \|\boldsymbol{\psi}_m - \widetilde{\boldsymbol{\psi}}^{\mathrm{pin}}\|$ denote the Euclidean distance between the pinching antenna and user $m$. The existence of a direct LoS path is determined by a Bernoulli random variable $\gamma_m \in \{0,1\}$ with
\begin{align}
    \mathbb{P}[\gamma_m = 1] = p_{\mathrm{LoS}}(r_m),
\end{align}
where $p_{\mathrm{LoS}}(r_m)$ is a distance-dependent LoS probability. Following the model in \cite{3gpp2020channel,bai2014analysis}, we use
\begin{subequations}
    \begin{align}
        p_{\mathrm{LoS}}(r_m) &= e^{-\beta r_m^2}, \\
        &= e^{-\beta [(x_m - \tilde x)^2 + C_m]},
    \end{align}
\end{subequations}
where $C_m \triangleq y_m^2 + d_v^2$, and $0 \leq \beta \leq 1$ is an environment-dependent parameter reflecting the density of blockages. In particular, a larger $\beta$ indicates a denser obstacle environment and hence a lower probability of maintaining a LoS path.

If a direct LoS path exists (i.e., $\gamma_m = 1$), the LoS channel coefficient is modeled as 
\begin{align}
    h_m^{\mathrm{LoS}} = \frac{\sqrt{\eta} e^{-j(\frac{2\pi}{\lambda} [(x_m - \tilde x)^2 + C_m]^{\frac{1}{2}} + \frac{2\pi}{\lambda_g}\tilde{x})} }{[(x_m - \tilde x)^2 + C_m]^{\frac{1}{2}}},
\end{align}
where $\lambda$ is the free-space wavelength, $\lambda_g$ is the guided wavelength in the waveguide, and $\eta = \frac{c^2}{(4\pi f_c)^2}$ is a frequency-dependent constant, with $c$ denoting the speed of light and $f_c$ denoting the carrier frequency. We note that the in-waveguide attenuation is omitted from the channel model, considering its limited impacts on the system performance, as evidenced by the analytical results in \cite{xu2025pinching,xu2025pinching-los}.

\subsubsection{NLoS Channel Component}
Regardless of the presence of a LoS path, the channel between the pinching antenna and user $m$ also includes an NLoS component, which captures the aggregate effect of multipath scattering in the environment. Following standard statistical models for mmWave propagation \cite{3gpp2020channel,poddar2023tutorial}, we express the NLoS channel as
\begin{align}
h_m^{\mathrm{NLoS}}(\tilde x) = \sum_{n=1}^{N_c} g_{m,n}(\tilde x),
\end{align}
where $N_c$ is the number of effective NLoS clusters and $g_{m,n}(\tilde x)$ denotes the small-scale fading gain of the $n$-th cluster. 
We model each cluster gain as an independent zero-mean circularly symmetric complex Gaussian random variable with distance-dependent variance:
\begin{align}
g_{m,n}(\tilde x) \sim \mathcal{CN} \left(0, \frac{\mu_{m,n}^2}{r_m^2(\tilde x)}\right),
\end{align}
where $\mu_{m,n}^2$ represents the normalized average power of the $n$-th NLoS cluster. Under this model, the average NLoS channel power becomes
\begin{align} \label{eqn: nlos channel power}
\mathbb{E}\bigl[|h_m^{\mathrm{NLoS}}(\tilde x)|^2\bigr]
= \frac{1}{r_m^2(\tilde x)} \sum_{n=1}^{N_c} \mu_{m,n}^2
\triangleq \frac{\mu_m^2}{r_m^2(\tilde x)},
\end{align}
where $\mu_m^2 \triangleq \sum_{n=1}^{N_c} \mu_{m,n}^2$ aggregates the effective NLoS cluster powers for user $m$. In this way, the NLoS component experiences the same large-scale distance-dependent attenuation as the LoS component, while the cluster powers ${\mu_{m,n}^2}$ themselves remain independent of the pinching-antenna position $\tilde x$ and can be specified according to the channel power delay profile or third generation partnership project (3GPP) recommendations.

\subsubsection{Composite Channel Model}

Accordingly, the overall channel coefficient between the pinching antenna at position $\tilde x$ and user $m$ can be written as
\begin{align}
    h_m(\tilde x) = \gamma_m h_m^{\mathrm{LoS}}(\tilde x) + h_m^{\mathrm{NLoS}}(\tilde x),
\end{align}
where $\gamma_m$, $h_m^{\mathrm{LoS}}(\tilde x)$, and $h_m^{\mathrm{NLoS}}(\tilde x)$ are mutually independent. This model reduces to a classic Rayleigh fading channel when $\gamma_m = 0$ (i.e., only NLoS propagation is present), and to a Rician-like channel when $\gamma_m = 1$, where both deterministic LoS and random NLoS components coexist. The relative strengths of the LoS and NLoS terms are determined by the propagation geometry through $r_m(\tilde x)$ and by the channel statistics $\eta$ and $\mu_m^2$, rather than by an explicit Rician $K$-factor.

\subsection{Signal Model and Performance Metrics}
Based on the system and channel models, we now present the signal model and define two key performance metrics that capture complementary aspects of system behavior under probabilistic LoS blockage and random NLoS scattering.

\subsubsection{Signal Model}
For user $m$, during its allocated time slot, the received signal is given by
\begin{align}
    y_m = \sqrt{P}\, h_m s_m + n_m,
\end{align}
where $P$ denotes the transmit power at the pinching antenna, $s_m$ is the information symbol intended for user $m$ satisfying $\mathbb{E}[|s_m|^2]=1$, and $n_m\sim\mathcal{CN}(0,\sigma_m^2)$ represents additive white Gaussian noise (AWGN) with power $\sigma_m^2$. 
The instantaneous received SNR at user $m$ is therefore
\begin{align} \label{eqn: inst snr}
    \Gamma_m  = \frac{P|h_m|^2}{\sigma_m^2}
     = \rho_m |h_m|^2,
\end{align}
where $\rho_m \triangleq P/\sigma_m^2$ denotes the transmit SNR scaling factor.

\subsubsection{Performance Metrics}
In realistic fading environments where both LoS blockage and NLoS scattering coexist, it is critical to assess system performance from two complementary perspectives: 
(i)~the long-term, throughput-oriented behavior captured by the \emph{average SNR}, and 
(ii)~the short-term, reliability-oriented behavior characterized by the \emph{outage-constrained SNR threshold}. 
These two metrics highlight distinct operational objectives and apply to different classes of wireless systems, as detailed below.
\begin{itemize}
    \item {\bf Average SNR-based Metric:} This metric focuses on the average received SNR across random channel realizations, aiming to ensure high overall spectral efficiency of users. Such an average-SNR-oriented formulation is particularly relevant for throughput-driven and long-term optimization scenarios, where the system operates over many independent fading realizations and seeks to maximize sustained network capacity. Typical examples include user-centric coverage enhancement, broadband wireless access, and network deployments in hotspot environments. In these cases, occasional short-term channel fluctuations are tolerable as long as the long-term service quality remains high, making the average SNR an appropriate and practically meaningful performance measure.
    \item {\bf Outage-Constrained Metric: } This metric emphasizes reliability by constraining the outage probability below a prescribed threshold. This reliability-oriented formulation is especially relevant for reliability-sensitive and mission-critical applications, such as autonomous driving, industrial control, and remote surgery, where instantaneous channel degradations or SNR drops can result in severe performance loss or even service interruption.  By explicitly accounting for the stochastic nature of LoS and NLoS conditions, the outage-constrained design seeks to guarantee a minimum SNR level with high probability, ensuring reliable communication even under adverse propagation conditions.
\end{itemize}

Taken together, the average-SNR-based and outage-constrained formulations provide a comprehensive evaluation framework for pinching-antenna systems, capturing long-term throughput efficiency and short-term reliability assurance. 
In the remainder of this paper, we investigate the pinching-antenna system design under these two performance metrics, respectively. Specifically, we first focus on the average-SNR-based formulation that characterizes the long-term efficiency, and then consider the outage-constrained formulation that emphasizes reliability under probabilistic channel variations.

\section{Average-SNR-based Pinching-Antenna System Design} \label{sec:avg_snr}
In this section, we focus on the average-SNR-based pinching-antenna system design. In particular, we first derive the per-user average SNR of the pinching-antenna system under probabilistic LoS and random NLoS channels. Then, based on the derived average SNR, we investigate a max-min average SNR maximization problem to capture both system spectral efficiency and user fairness.

\subsection{Per-User Average SNR Derivations}

Based on the instantaneous SNR in \eqref{eqn: inst snr}, the average received SNR at user $m$ is given by
\begin{align}
    \bar \Gamma_m(\tilde x) 
        &= \mathbb{E}_{\gamma_m, h_m^{\mathrm{NLoS}}} \left[ \Gamma_m(\tilde x)\right] \notag\\
        &= \mathbb{E}_{\gamma_m, h_m^{\mathrm{NLoS}}} \left[ \rho_m |h_m(\tilde x)|^2 \right] \notag\\
        &= \rho_m \mathbb{E}_{\gamma_m, h_m^{\mathrm{NLoS}}} \left[ \big|\gamma_m h_m^{\mathrm{LoS}}(\tilde x) + h_m^{\mathrm{NLoS}}(\tilde x)\big|^2 \right] \notag\\
        &\overset{(a)}{=} \rho_m \left( \mathbb{E}_{\gamma_m} \left[ \big|\gamma_m h_m^{\mathrm{LoS}}(\tilde x)\big|^2 \right]
        + \mathbb{E}_{h_m^{\mathrm{NLoS}}} \left[\big|h_m^{\mathrm{NLoS}}(\tilde x)\big|^2 \right] \right) \notag\\
        &\overset{(b)}{=} \rho_m \left(\mathbb{E}[\gamma_m] \big|h_m^{\mathrm{LoS}}(\tilde x)\big|^2 + \mathbb{E}_{h_m^{\mathrm{NLoS}}} \left[\big|h_m^{\mathrm{NLoS}}(\tilde x)\big|^2 \right] \right) \notag\\
        &\overset{(c)}{=} \rho_m \left( e^{-\beta r_m^2(\tilde x)} \big|h_m^{\mathrm{LoS}}(\tilde x)\big|^2 
        + \frac{\mu_m^2}{r_m^2(\tilde x)} \right) \notag \\
        &\overset{(d)}{=} \rho_m \left( \frac{\eta e^{-\beta r_m^2(\tilde x)}}{r_m^2(\tilde x)} 
        + \frac{\mu_m^2}{r_m^2(\tilde x)} \right) \notag\\
        &= \rho_m \frac{\eta e^{-\beta r_m^2(\tilde x)} + \mu_m^2}{r_m^2(\tilde x)}, \label{eq:average_snr_position}
\end{align}
The key steps in the derivation of \eqref{eq:average_snr_position} are explained as follows: Step (a) follows from the independence between $\gamma_m$ and $h_m^{\mathrm{NLoS}}(\tilde x)$, and from the assumption that the NLoS component has zero mean, which causes the cross-term to vanish when expanding the squared magnitude. Step (b) utilizes the fact that $\gamma_m \in \{0, 1\}$, so that $\mathbb{E}[|\gamma_m h_m^{\mathrm{LoS}}(\tilde x)|^2] = \mathbb{E}[\gamma_m] |h_m^{\mathrm{LoS}}(\tilde x)|^2$. Step (c) applies the probabilistic LoS model, where the probability of LoS occurrence is $\mathbb{E}[\gamma_m] = p_{\mathrm{LoS}}(r_m(\tilde x)) = e^{-\beta r_m^2(\tilde x)}$, and uses the distance-dependent average NLoS power $\mathbb{E}[|h_m^{\mathrm{NLoS}}(\tilde x)|^2] = \mu_m^2 / r_m^2(\tilde x)$ derived in \eqref{eqn: nlos channel power}. Step (d) substitutes the deterministic LoS channel gain model $|h_m^{\mathrm{LoS}}(\tilde x)|^2 = \eta / r_m^2(\tilde x)$ into the expression.

The closed-form expression in \eqref{eq:average_snr_position} clearly reveals how the pinching-antenna position $\tilde x$ and the LoS blockage coefficient $\beta$ jointly affect the user's average received SNR. The factor $1 / r_m^2(\tilde x)$ reflects the inverse-square decay of the received power with the user-to-antenna distance: as $r_m(\tilde x)$ increases, the free-space path loss grows proportionally to $r_m^2(\tilde x)$ and, consequently, the channel gain and the average SNR decrease. The exponential factor $e^{-\beta r_m^2(\tilde x)}$ further models the distance-dependent LoS probability, which decreases with $r_m(\tilde x)$ and thus attenuates the effective LoS contribution. Meanwhile, the term $\mu_m^2 / r_m^2(\tilde x)$ represents the distance-dependent average power of the NLoS component, which becomes dominant in scenarios with severe LoS blockage (e.g., large $\beta$ or large $r_m(\tilde x)$).

\subsection{Max--Min Average SNR Maximization Problem Formulation and Analysis}

Our goal is to maximize the minimum average SNR across all users by optimizing the antenna placement. Based on the considered OMA transmission scheme and the average SNR expression in \eqref{eq:average_snr_position}, the max--min average SNR maximization problem can be formulated as
\begin{subequations} \label{eq:orig_problem}
\begin{align}
    \max_{\,\tilde x\,} \quad & \min_{1\le m\le M} \bar\Gamma_m(\tilde x) \\
    \st \quad & 0 \le \tilde x \le D_x.
\end{align}
\end{subequations}

To solve this non-smooth optimization, we introduce an auxiliary variable $ t $ representing the common guaranteed SNR level, leading to its equivalent epigraph form:
\begin{subequations} \label{p: max min epigraph}
\begin{align} 
    \max_{\tilde x,\, t} \quad & t \\
    \st \quad & \bar\Gamma_m(\tilde x) \ge t, \quad \forall m \in \mathcal{M}, \label{eq:epigraph_constraint} \\
             & 0 \le \tilde x \le D_x.
\end{align}
\end{subequations}
Although \eqref{p: max min epigraph} is not jointly convex in $(\tilde x,t)$, we next show that it can still be globally solved in a low-complexity manner. To this end, let us first present the following proposition.

\begin{prop}
    \label{prop:feasible_interval_new}
    For each user $m$ and any given target SNR level $t>0$, the feasibility set
    \begin{align}
        \mathcal{I}_m(t) \triangleq \left\{\tilde x \in [0,D_x] : \bar\Gamma_m(\tilde x) \ge t \right\}
    \end{align}
    is either a closed interval or an empty set. Consequently, the global feasibility set
    \begin{align}
        \mathcal{F}(t) \triangleq \bigcap_{m=1}^M \mathcal{I}_m(t)
    \end{align}
    is also either a closed interval or empty. Moreover, $\mathcal{F}(t)$ monotonically shrinks as $t$ increases, i.e.,
    \begin{align}
        t_1 < t_2 \quad \Rightarrow \quad \mathcal{F}(t_1) \supseteq \mathcal{F}(t_2).
    \end{align}
\end{prop}

\emph{Proof:} 
First, let us define the scalar function
\begin{align} \label{eqn: f definition}
    f_m(y) \triangleq \rho_m \frac{\eta e^{-\beta y} + \mu_m^2}{y}, \quad y>0.
\end{align}
Thus, for user $m$, we can write $\bar\Gamma_m(\tilde x) = f_m(r_m^2(\tilde x))$. Next, we analyze the scalar function $f_m(y)$ for $y>0$. Its derivative is given by
\begin{align} 
    f_m'(y) 
    &= \rho_m \frac{-\eta \beta e^{-\beta y} y - \eta e^{-\beta y} - \mu_m^2}{y^2} \notag\\
    &= -\rho_m \frac{\eta e^{-\beta y} (\beta y + 1) + \mu_m^2}{y^2} < 0,
\end{align}
which shows that $f_m(y)$ is strictly decreasing and continuous on $(0,\infty)$. Hence, for any $\tau$ in the range of $f_m(\cdot)$, the inequality $f_m(y) \ge t$ is equivalent to $y \le \alpha_m(t)$ for some threshold $\alpha_m(t)>0$, determined uniquely as the solution of $f_m(\alpha_m(t)) = t$.

For a given target SNR level $t>0$, the per-user constraint $\bar\Gamma_m(\tilde x) \ge t$ is thus equivalent to
\begin{align} \label{eqn: distance bound}
    f_m(r_m^2(\tilde x)) \ge t
    \quad \Longleftrightarrow \quad
    r_m^2(\tilde x) \le \alpha_m(t),
\end{align}
whenever $t$ lies within the achievable SNR range of user $m$. The function $r_m^2(\tilde x) = (x_m - \tilde x)^2 + C_m$ is a convex quadratic function of $\tilde x$, and the sublevel set $\{\tilde x : r_m^2(\tilde x) \le \alpha_m(t)\}$ is therefore a convex set in $\tilde x$, which is readily seen to be a closed interval. 
In particular, defining
\begin{align}
    d_m(t) \triangleq \sqrt{\max\{\alpha_m(t) - C_m,\ 0\}},
\end{align}
and intersecting this interval with the deployment range, the corresponding feasibility interval for user $m$ is given by
\begin{align} \label{eqn: per user feasibility set}
    \mathcal{I}_m(t)
    = \left[x_m - d_m(t),\, x_m + d_m(t)\right] \cap [0,D_x],
\end{align}
which is either a closed interval or empty.

The global feasibility set $\mathcal{F}(t) = \bigcap_{m=1}^M \mathcal{I}_m(t)$ is formed by the intersection of finitely many closed intervals within the compact domain $[0,D_x]$, and is thus itself either a closed interval or empty.

Finally, since $f_m(y)$ is strictly decreasing in $y$ and $r_m^2(\tilde x)>0$, increasing $t$ tightens the inequality $f_m(r_m^2(\tilde x)) \ge t$, which corresponds to a smaller upper bound $\alpha_m(t)$ on $r_m^2(\tilde x)$. Hence $\mathcal{I}_m(t)$ shrinks as $t$ increases, and so does their intersection $\mathcal{F}(t)$, establishing the nested-set property stated in the proposition. \hfill$\blacksquare$

\subsection{Solving Problem \eqref{p: max min epigraph} via the Monotonicity of $\mathcal{F}(t)$}
Next, we show how to use Proposition~\ref{prop:feasible_interval_new} to solve problem \eqref{p: max min epigraph} efficiently.
Proposition \ref{prop:feasible_interval_new} shows that for any $t$, the feasibility set of \eqref{eq:epigraph_constraint} is a (possibly empty) interval in $\tilde x$, and that these intervals are nested with respect to $t$. As a result, the original max–min average SNR maximization problem in \eqref{eq:orig_problem} is quasi-convex in $\tilde x$ and admits a globally optimal solution that can be efficiently obtained via a bisection search over $t$.

The key step in each bisection iteration is the feasibility check for a given $t$. 
To this end, recall from the proof of Proposition \ref{prop:feasible_interval_new} that the per-user average SNR constraint
$\bar\Gamma_m(\tilde x) \ge t$
is equivalent to an upper bound on the squared distance,
\begin{align}
    r_m^2(\tilde x) \le \alpha_m(t),  \label{eqn:distance_bound}
\end{align}
where $\alpha_m(t)>0$ is uniquely determined by the scalar equation
\begin{align}
    f_m(\alpha_m(t)) = t,  \label{eqn:alpha_equation}
\end{align}
with $f_m(\cdot)$ defined in \eqref{eqn: f definition}. Since $f_m(\cdot)$ is strictly decreasing and continuous, \eqref{eqn:alpha_equation} has a unique solution for any $t$ within the achievable SNR range of user $m$, and this solution can be efficiently obtained via a one-dimensional bisection search.

Before starting the outer bisection over $t$, we precompute, for each user $m$, the minimum and maximum squared distances
$y_{m,\min}$ and $y_{m,\max}$ over $\tilde x \in [0,D_x]$, and the corresponding maximum achievable average SNR
\begin{align}
    \bar\Gamma_m^{\max} = f_m(y_{m,\min}).
\end{align}
These quantities remain fixed throughout the algorithm. We then choose bounds $t_{\min}$ and $t_{\max}$ such that $\mathcal{F}(t_{\min}) \neq \emptyset$ (e.g., $t_{\min}=0$) and $t_{\max} > \max_m \bar\Gamma_m^{\max}$, which guarantees $\mathcal{F}(t_{\max}) = \emptyset$.

At each bisection step, we set $t = (t_{\min}+t_{\max})/2$ and perform the following scalar operations:
\begin{itemize}
    \item For each user $m$, if $t > \bar\Gamma_m^{\max}$, then the equation $f_m(y) = t$ has no solution on $[y_{m,\min}, y_{m,\max}]$, so $\mathcal{I}_m(t) = \emptyset$ and hence $\mathcal{F}(t) = \emptyset$. In this case, $t$ is immediately declared infeasible.
    \item Otherwise, for each user $m$, solve the one-dimensional equation $f_m(y) = t$ over $y \in [y_{m,\min}, y_{m,\max}]$ to obtain $\alpha_m(t)$, construct $\mathcal{I}_m(t)$, and update $\mathcal{F}(t) = \bigcap_{m=1}^M \mathcal{I}_m(t)$.
\end{itemize}
If the resulting $\mathcal{F}(t)$ is non-empty, we update $t_{\min} \leftarrow t$; otherwise, we update $t_{\max} \leftarrow t$. By the nested-set property of $\mathcal{F}(t)$, this bisection search converges to the optimal guaranteed SNR $t^*$ with logarithmic complexity in the target accuracy. Finally, any $\tilde x^* \in \mathcal{F}(t^*)$ is an optimal pinching-antenna position for problem~\eqref{eq:orig_problem}.

\begin{rem}
    The overall computational complexity is dominated by the double-loop bisection searches over the scalar variables $t$ and $y$. The outer bisection over $t$ requires on the order of $\log(1/\epsilon_t)$ iterations to achieve accuracy $\epsilon_t$ in the guaranteed SNR level, while each inner bisection for solving $f_m(y) = t$ needs on the order of $\log(1/\epsilon_y)$ iterations to reach accuracy $\epsilon_y$. In each outer iteration, at most $M$ such one-dimensional root-finding problems are solved, and each feasibility interval $\mathcal{I}_m(t)$ is obtained via simple scalar operations. Consequently, the total complexity scales as $\mathcal{O}\big(M \log(1/\epsilon_t)\log(1/\epsilon_y)\big)$, involving only scalar computations, which makes the proposed max-min average SNR design highly efficient and scalable with respect to the number of users $M$.
\end{rem}

\subsection{Optimal Closed-Form Solution in a Two-User Case}

In this subsection, we consider a special case with two users and assume that
$\sigma_1^2 = \sigma_2^2 = \sigma^2$ and $\mu_1^2 = \mu_2^2 = \mu^2$. Hence, the
per-user SNR coefficients satisfy $\rho_1 = \rho_2 = \rho \triangleq P/\sigma^2$.
Without loss of generality, we assume that $x_1 \leq x_2$ and let
$\Delta = x_2 - x_1 \ge 0$ denote the horizontal distance between the users.
Then, the problem in \eqref{p: max min epigraph} reduces to maximizing the
minimum average SNR across the two users by optimizing the pinching-antenna
position $\tilde x$ and the auxiliary SNR variable $t$.
For notational simplicity, we
denote $\alpha \triangleq \alpha(t)$ in the following.
Then, the globally optimal solutions for the two-user
max–min SNR problem are given by the following lemma.

\begin{lem}\label{lem: two-user optimal solution_new}
Let $C_{\max}=\max\{C_1,C_2\}$ and $C_{\min}=\min\{C_1,C_2\}$. 
Then, the minimal feasible $\alpha^*$ and the optimal pinching-antenna position
$\tilde x^*$ are given in closed form by
\begin{small}
    \begin{align}
        \alpha^* =
        \begin{cases}
        \displaystyle C_{\max}, &
        \text{if } \Delta \!\leq\! \sqrt{C_{\max}\!-\!C_{\min}},\\[1mm]
        \displaystyle \frac{\Delta^2}{4} \!+\! \frac{C_1+C_2}{2} \!+\! \frac{(C_1-C_2)^2}{4\,\Delta^2}, &
        \text{if } \Delta\! >\! \sqrt{C_{\max}\!-\! C_{\min}},
        \end{cases} \label{eq:alpha_two_regimes}
    \end{align}
\end{small}
and
\begin{small}
    \begin{align}
        \tilde x^* =
        \begin{cases}
        x_2, & \text{if } C_2\ge C_1~ \mathrm{and}~ \Delta \le \sqrt{C_2-C_1},\\
        x_1, & \text{if } C_1> C_2 ~\mathrm{and}~ \Delta \le \sqrt{C_1-C_2},\\[1mm]
        \displaystyle \frac{x_1 \!+\!x_2}{2} \!+\! \frac{C_2\!-\!C_1}{2\Delta}, &
        \text{if } \Delta > \sqrt{C_{\max}-C_{\min}}.
        \end{cases}
        \label{eq:xstar_two_regimes_new}
    \end{align}
\end{small}
The degenerate case $\Delta=0$ is included in the first line of
\eqref{eq:alpha_two_regimes}, yielding $\alpha^*=C_{\max}$ and
$\tilde x^*=x_1=x_2$. Besides, given $\alpha^*$, the optimal target max-min average SNR is given by
\begin{align}
    t^* = f(\alpha^*)
    = \rho \frac{\eta e^{-\beta \alpha^*} + \mu^2}{\alpha^*}. \label{eq:tstar_two_user_new}
\end{align}
\end{lem}

\emph{Proof:}
The objective is to maximize the worst-user SNR level $t$, which, due to the
strict monotonicity of $f(\cdot)$, is equivalent to minimizing the auxiliary
variable $\alpha$ subject to feasibility. Hence, we fix an auxiliary variable
$\alpha > 0$ and check whether the per-user constraints are jointly satisfied.

For user $m$, the feasibility requirement $r_m^2(\tilde x) \le \alpha$ can be
written as
\begin{align}
|x_m - \tilde{x}| \le \sqrt{\alpha - C_m} \triangleq d_m, \quad m \in \{1,2\},
\end{align}
which implies that the pinching-antenna location $\tilde{x}$ must lie in the
interval
\begin{align}
\tilde{x} \in [x_1 - d_1,\, x_1 + d_1] \cap [x_2 - d_2,\, x_2 + d_2].
\end{align}
Thus, the intersection of these two intervals must be non-empty for $\alpha$ to
be feasible.

The two intervals overlap if and only if
\begin{align}
    \Delta  = x_2- x_1
           &\leq d_1 + d_2 \notag \\
           &= \sqrt{\alpha - C_1} + \sqrt{\alpha - C_2}
            \triangleq \Psi(\alpha). \label{eq:Psi_condition_alpha}
\end{align}
Since $\Psi(\alpha)$ is continuous and strictly increasing for
$\alpha \ge C_{\max}$, the minimal feasible $\alpha^*$ is the smallest
$\alpha \ge C_{\max}$ satisfying \eqref{eq:Psi_condition_alpha}. Two regimes
arise:

\textbf{Case 1:} If $\Delta \leq \Psi(C_{\max})=\sqrt{C_{\max}-C_{\min}}$, then
$\alpha=C_{\max}$ already satisfies \eqref{eq:Psi_condition_alpha}, so the
minimal feasible value is $\alpha^*=C_{\max}$. Suppose, without loss of
generality, that $C_{\max}=C_2$. Then, the second interval collapses to the
singleton $\{x_2\}$. Because
$\Delta \leq \sqrt{C_2-C_1}$, we have
$x_2 \in [x_1-\sqrt{C_2-C_1}, x_1+\sqrt{C_2-C_1}]$, so the intersection is the
singleton $\{x_2\}$ and thus $\tilde x^*=x_2$. The case $C_{\max}=C_1$ is
symmetric and yields $\tilde x^*=x_1$. The degenerate case $\Delta=0$ is
included with $\alpha^*=C_{\max}$ and $\tilde x^*=x_1=x_2$.

\textbf{Case 2:} If $\Delta > \Psi(C_{\max})$, the minimal feasible $\alpha^*$
strictly exceeds $C_{\max}$ and satisfies the overlap condition:
\begin{align}
\sqrt{\alpha^*-C_1}+\sqrt{\alpha^*-C_2}=\Delta. \label{eq:touch_eq_alpha}
\end{align}
Define $y_1 \triangleq \sqrt{\alpha^*-C_1}$ and
$y_2 \triangleq \sqrt{\alpha^*-C_2}$. Then,
\begin{align}
    y_1 + y_2 &= \Delta, \label{eq:sum_diff_1_new}\\
    y_1^2 - y_2^2 &= C_2 - C_1. \label{eq:sum_diff_2_new}
\end{align}
Combining \eqref{eq:sum_diff_1_new} and \eqref{eq:sum_diff_2_new} yields
\begin{align} \label{eq:sum_diff_3_new}
    y_1 - y_2 = \frac{C_2 - C_1}{\Delta}.
\end{align}
Solving this linear system gives
\begin{subequations}\label{eq:y1y2_closed_new}
    \begin{align}
        y_1 &= \frac{\Delta}{2} + \frac{C_2-C_1}{2\Delta},\\
        y_2 &= \frac{\Delta}{2} - \frac{C_2-C_1}{2\Delta}. 
    \end{align}  
\end{subequations}
Substituting $y_1^2 = \alpha^*-C_1$ (or $y_2^2 = \alpha^*-C_2$), we obtain
\begin{subequations}
    \begin{align}
        \alpha^* &= C_1 + y_1^2 \\
         &= C_1 + \left(\frac{\Delta}{2} + \frac{C_2-C_1}{2\Delta}\right)^{\!2} \\
         &= \frac{\Delta^2}{4} + \frac{C_1+C_2}{2} + \frac{(C_1-C_2)^2}{4 \Delta^2},
    \end{align}
\end{subequations}
which coincides with \eqref{eq:alpha_two_regimes}. The optimal pinching-antenna
position is then
\begin{align}
\tilde x^*
= x_1+y_1
= \frac{x_1+x_2}{2}+\frac{C_2-C_1}{2\Delta}, \label{eq:xstar_touch_alpha}
\end{align}
which matches the third line of \eqref{eq:xstar_two_regimes_new}.

\begin{figure}[!t]
	\centering
	\includegraphics[width=0.88\linewidth]{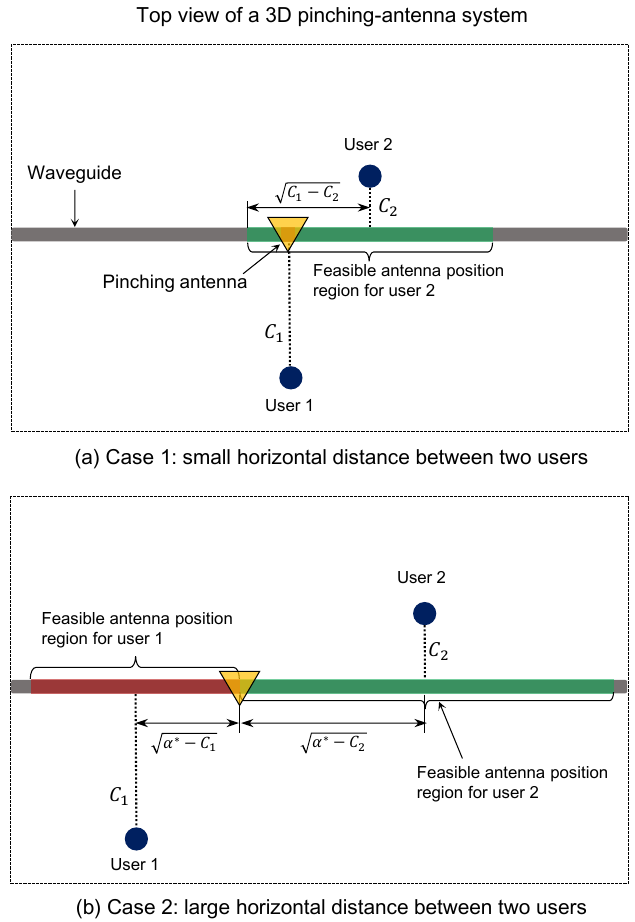}\\
        \captionsetup{justification=justified, singlelinecheck=false, font=small}	
        \caption{A geometric illustration of a two-user pinching-antenna system under (a) small and (b) large horizontal distances, showing the feasible antenna position regions and optimal deployment strategy.} \label{fig: geo illustration} 
\end{figure} 

Finally, substituting $\alpha^*$ into \eqref{eq:tstar_two_user_new} yields the
optimal worst-user average SNR $t^*$. The proof of
Lemma~\ref{lem: two-user optimal solution_new} is complete.
\hfill$\blacksquare$

\begin{rem}
   The closed-form solution in Lemma \ref{lem: two-user optimal solution_new}
   offers several insightful implications. When the horizontal distance between
   the two users is small, such that
   $\Delta \leq \sqrt{C_{\max}-C_{\min}}$, the optimal strategy is to place the
   pinching antenna directly at the limiting user who has the larger offset
   $C_m$, thereby maximizing the worst-case SNR. When the horizontal distance
   between the two users is large, the optimal $\tilde x^*$ lies strictly
   between them. However, it is not the simple midpoint; instead, it is a
   biased midpoint shifted toward the user with the larger offset by an amount
   of $(C_2-C_1)/(2\Delta)$. This bias compensates for the increased propagation
   loss and LoS blockage experienced by the user with higher $C_m$, thus
   balancing the SNR performance across both users. To facilitate understanding,
   a geometric illustration of the optimal pinching-antenna placement for the
   two-user case is presented in Fig.~\ref{fig: geo illustration}.
\end{rem}

\section{Outage-Constrained Pinching-Antenna System Design} \label{sec:outage}

In this section, we focus on the outage-constrained performance metric and investigate the problem of maximizing the SNR threshold that guarantees a target reliability level across all users under probabilistic LoS and random NLoS channels. 

\subsection{Outage-Constrained Problem Formulation}

For a given SNR threshold $t$ and a target outage probability $\varepsilon_m \in (0,1)$, the reliability requirement of user $m$ is defined as
\begin{align}
    \Pr\!\big(\Gamma_m(\tilde x) \ge t \big) \ \ge\ 1 - \varepsilon_m,
    \label{eq:outage_constraint}
\end{align}
where $\Gamma_m(\tilde x) = \rho_m |h_m(\tilde x)|^2$ denotes the instantaneous received SNR at user $m$ when the pinching antenna is located at $\tilde x$.
This condition ensures that user $m$ experiences an SNR above the target threshold $t$ with probability at least $1-\varepsilon_m$, i.e., an outage probability below $\varepsilon_m$.

The design objective is to maximize the achievable SNR threshold $t$ that satisfies the outage constraints of all users by optimally positioning the pinching antenna along the waveguide.
Mathematically, the problem can be formulated as
\begin{subequations}\label{prob:outage_opt}
\begin{align}
    \max_{\tilde x,\, t} \quad & t  \\
    \text{s.t.} \quad
    & \Pr\!\big(\Gamma_m(\tilde x) \ge t \big) \ \ge\ 1 - \varepsilon_m,\quad \forall m \in \Mset, \label{eqn:outage_constraint_each}\\
    & 0 \le \tilde x \le D_x.
\end{align}
\end{subequations}

Problem \eqref{prob:outage_opt} aims to obtain the optimal pinching-antenna position $\tilde x^\star$ that maximizes the guaranteed SNR threshold $t^\star$ while satisfying a prescribed reliability requirement for all users. 
To make this problem tractable, we first derive a closed-form expression for the CCDF of $\Gamma_m(\tilde x)$, which quantifies the probability that the instantaneous SNR exceeds a given threshold. 
The CCDF characterization is formally presented in the following proposition.

\begin{prop}\label{prop:CCDF_new}
Let $r_m^2(\tilde x) = (x_m - \tilde x)^2 + C_m$ denote the squared distance between user $m$ and the pinching antenna. Define
\[
a_m \triangleq 
\sqrt{2}\,\frac{\sqrt{\eta}}{\mu_m},
\qquad
b_m\big(r_m(\tilde x),t\big) \triangleq 
\sqrt{2}\,\frac{r_m(\tilde x)\sqrt{t / \rho_m}}{\mu_m}.
\]
Then, for any threshold $t \ge 0$, the CCDF of $\Gamma_m(\tilde x)$ is given by
\begin{align}
    \Pr\!\big(\Gamma_m(\tilde x) \ge t\big)
    &= e^{-\beta r_m^2(\tilde x)}\, 
       Q_1\!\big(a_m, b_m(r_m(\tilde x),t)\big) \notag\\
    &\quad + \big(1 \!-\! e^{-\beta r_m^2(\tilde x)}\big)\,
       \exp\!\Big(\!-\frac{t\,r_m^2(\tilde x)}{\rho_m\mu_m^2}\Big),
    \label{eq:CCDF_final_full_new}
\end{align}
where $Q_1(a,b)$ denotes the first-order Marcum-$Q$ function,
\[
Q_1(a,b)
= \int_b^{\infty} x\, 
\exp\!\Big(-\frac{x^2 + a^2}{2}\Big)
I_0(ax)\, dx, \quad a,b \ge 0,
\]
and $I_0(\cdot)$ is the modified Bessel function of the first kind (order zero). 
Moreover, for any fixed $t>0$, $\Pr(\Gamma_m(\tilde x) \ge t)$ is a strictly decreasing function of the user-to-antenna distance $r_m(\tilde x)$.
\end{prop}

\emph{Proof:} See Appendix \ref{app: ccdf_new}.  \hfill$\blacksquare$

\subsection{Solving Problem \eqref{prob:outage_opt} via CCDF Monotonicity}
In this subsection, we show how to efficiently determine the optimal pinching-antenna position by exploiting the monotonicity property of the CCDF derived in \eqref{eq:CCDF_final_full_new}. 
For each user $m$ and any fixed SNR threshold $t$, the reliability constraint 
$\Pr \big(\Gamma_m(\tilde x)\ge t\big) \ge 1-\varepsilon_m$
can be equivalently expressed as an upper bound on the squared distance term $r_m^2(\tilde x)$:
\begin{align}\label{eq:Um_outage_def_new}
    r_m^2(\tilde x)\ \le\ U_m(t),
\end{align}
where $U_m(t)$ is the solution of the scalar monotone equation
\begin{align}
&e^{-\beta y}\,Q_1 \Big(a_m,b_m(\sqrt{y},t)\Big) \notag\\
&\quad + \Big(1-e^{-\beta y}\Big)\,\exp\!\Big(-\frac{t\,y}{\rho_m\mu_m^2}\Big)
= 1-\varepsilon_m, \ \textrm{for} \ y>0, \label{eq:Um_scalar_eq_new}
\end{align}
with $a_m$ and $b_m(\cdot,t)$ defined in Proposition~\ref{prop:CCDF_new}.

For any fixed threshold $t$, the left-hand side (LHS) of \eqref{eq:Um_scalar_eq_new} is a continuous and strictly decreasing function of $y = r_m^2(\tilde x)$, as established by the monotonicity result in Proposition~\ref{prop:CCDF_new}. 
This ensures that \eqref{eq:Um_scalar_eq_new} admits at most one solution $U_m(t)$; when such a solution exists, it uniquely specifies the maximum allowable squared user-to-antenna distance that still satisfies the outage constraint for user $m$. 
If no solution exists within the physically relevant range of $y$, the corresponding outage constraint becomes infeasible at the given threshold $t$.

\begin{table*}[!t]
	\centering
	\caption{Summary of average-SNR-based and outage-constrained designs.}
	\renewcommand{\arraystretch}{1.5}
	\begin{tabularx}
		{\linewidth}
		{ >{\raggedright\arraybackslash}p{2.9cm}
			>{\raggedright\arraybackslash}X 
			>{\raggedright\arraybackslash}X}
		\hline \hline
		\textbf{Aspect} & \textbf{Average-SNR-based design} & \textbf{Outage-constrained design} \\
		\hline
		
		\textbf{Objective} 
		& Maximize the minimum average received SNR among users
		& Maximize the guaranteed SNR threshold subject to per-user outage constraints \\
		\hline
		
		\textbf{Algorithmic structure} 
		& Double-loop bisection: outer search on $t$; inner search for $\alpha_m(t)$ by solving the scalar monotone equation \eqref{eqn:alpha_equation}
		& Double-loop bisection: outer search on $t$; inner search for $U_m(t)$ by solving the monotone CCDF equation~\eqref{eq:Um_scalar_eq_new} \\
		\hline
		
		\textbf{Optimality} 
		& Globally optimal 
		& Globally optimal \\
		\hline
		
		\textbf{Computational complexity order} 
		& $\mathcal{O} \big(M\log(1/\epsilon_t)\log(1/\epsilon_y)\big)$, where $\epsilon_t$ and $\epsilon_y$ are accuracies of outer and inner bisection searches, respectively 
		& $\mathcal{O} \big(M\log(1/\epsilon_t)\log(1/\epsilon_u)\big)$, where $\epsilon_t$ and $\epsilon_u$ are accuracies of outer and inner bisection searches, respectively \\
		\hline
		
		\textbf{Reliability interpretation} 
		& Reflects \emph{long-term fairness and efficiency} by optimizing the users' average SNRs 
		& Ensures \emph{short-term reliability} by enforcing per-user outage constraints under random LoS/NLoS fading \\
		\hline
		
		\textbf{Application scenario} 
		& Efficiency-oriented or throughput-fair systems (e.g., wide-area coverage, access networks) 
		& Reliability-critical and latency-sensitive applications (e.g., industrial control, vehicular links) \\
		\hline \hline
	\end{tabularx}
	\label{tab:comparison_snr_outage}
\end{table*}

The monotonicity and continuity of the LHS in \eqref{eq:Um_scalar_eq_new} naturally allow us to use an efficient bisection search for determining $U_m(t)$. 
Specifically, since the LHS decreases smoothly from one to zero as $y$ increases, the unique root can be located by iteratively halving the search interval until convergence. 
This approach is numerically stable, simple to implement, and guarantees convergence to the desired accuracy with logarithmic complexity.  
Given $U_m(t)$, \eqref{eq:Um_outage_def_new} yields a closed interval constraint on the antenna position:
\begin{align}\label{eq:Im_outage_new}
    \tilde x  \in \mathcal{J}_m(t) \triangleq \big[ x_m - d_m(t),\, x_m + d_m(t) \big] \cap [0,D_x],
\end{align}
where $d_m(t) \triangleq \sqrt{\max\{U_m(t)-C_m,\,0\}}$.  
Therefore, the common feasible set at level $t$ is the intersection
\begin{align}\label{eq:Ft_outage_new}
    \mathcal{T}(t) \triangleq\ \bigcap_{m=1}^M \mathcal{J}_m(t),
\end{align}
which is either a closed interval or empty.

Moreover, for any fixed distance $r_m(\tilde x)$, the CCDF $\Pr(\Gamma_m(\tilde x)\ge t)$ is strictly decreasing in $t$, since a larger SNR threshold is harder to satisfy. 
Consequently, to maintain $\Pr(\Gamma_m(\tilde x)\ge t)\ge 1-\varepsilon_m$ as $t$ increases, the maximum allowable distance must shrink, i.e., $U_m(t)$ is nonincreasing in $t$. 
As a result, each $\mathcal{J}_m(t)$ shrinks as $t$ increases, and the family $\{\mathcal{T}(t)\}_{t\ge 0}$ is nested and shrinking:
\begin{align}
    t_1<t_2 \ \Longrightarrow\ \mathcal{T}(t_1)\supseteq\mathcal{T}(t_2).
\end{align}
This quasi-convex feasibility structure enables a global solution via a one-dimensional bisection search on $t$.

As a result, problem \eqref{prob:outage_opt} can be globally solved using a double-loop bisection search-based algorithm. 
Specifically, the outer loop performs a bisection search over the SNR threshold $t$ to identify the maximum feasible value, while the inner loop determines, for each user, the corresponding upper bound $U_m(t)$ by solving the scalar monotone equation in \eqref{eq:Um_scalar_eq_new}. 

\begin{rem}
    The computational complexity of the proposed algorithm is determined by the double-loop bisection search. 
    The outer loop performs bisection over the SNR threshold $t$ with complexity on the order of $\log(1/\epsilon_t)$, where $\epsilon_t$ denotes the accuracy in the SNR level, while the inner loop computes each per-user boundary $U_m(t)$ with complexity on the order of $\log(1/\epsilon_u)$, where $\epsilon_u$ denotes the accuracy in solving \eqref{eq:Um_scalar_eq_new}. 
    Overall, the total complexity scales as $\mathcal{O} \big(M\log(1/\epsilon_t)\log(1/\epsilon_u)\big)$. 
    Since each step involves only exponential and Marcum-$Q$ function evaluations and one-dimensional bisection searches, the proposed approach is both computationally efficient and numerically stable, making it well suited for real-time implementation in practical pinching-antenna systems.
\end{rem}

To highlight the distinctions and complementarity between the two formulations, 
Table \ref{tab:comparison_snr_outage} summarizes their main features in terms of 
objective, reliability interpretation, algorithmic structure, complexity, and 
application focus. Both formulations build on the same geometric structure of the 
pinching-antenna system and the same random LoS and NLoS channel model, and 
both are globally solvable via low-complexity one-dimensional bisection schemes. 
The average-SNR-based design focuses on improving the \emph{long-term} received 
signal quality and fairness across users, whereas the outage-constrained 
formulation explicitly enforces a target \emph{short-term} reliability level 
under random LoS blockage and NLoS fading. 
Together, they provide a unified design framework for balancing 
\emph{efficiency} and \emph{reliability} in flexible pinching-antenna-enabled 
communication systems.

\section{Simulation Results} \label{sec:simulation}
The performance of the proposed pinching-antenna system is evaluated through comprehensive numerical simulations. The key simulation parameters are summarized in Table \ref{tab:sim-param}. Unless otherwise specified, each figure uses the default parameter values listed in the table, while system-level parameters such as the communication-region side length $D_x$, the blockage coefficient $\beta$, and the transmit power $P$ are varied to illustrate different aspects of system behavior.

\begin{table}[!t]
\centering
\caption{Simulation parameters.}
\renewcommand{\arraystretch}{1.4}
\begin{tabularx}{7.5cm}{>{\raggedright\arraybackslash}p{5.1cm} >{\raggedright\arraybackslash}p{2.5cm}}
\hline \hline
\textbf{Parameter}              & \textbf{Value}           \\
\hline
Carrier frequency ($f_c$) & $28$ GHz \\
Waveguide height ($d_v$) & $10$ m \\
Transmit power ($P$) & $40$ dBm \\
Noise power ($\sigma^2$) & $-90$ dBm \\
Number of users ($M$) & $4$ \\
Convergence threshold ($\epsilon$) & $10^{-3}$ \\
NLoS channel power ($\mu_m^2$) & $-60$ dBm \\
Communication region width ($D_y$) & $10$ m \\
\hline \hline
\end{tabularx}
\label{tab:sim-param}
\end{table}

\begin{figure}[!t]
  \centering 
  \subfloat[Max–min average SNR of pinching and fixed antennas.]{
    \includegraphics[width=0.82\linewidth]{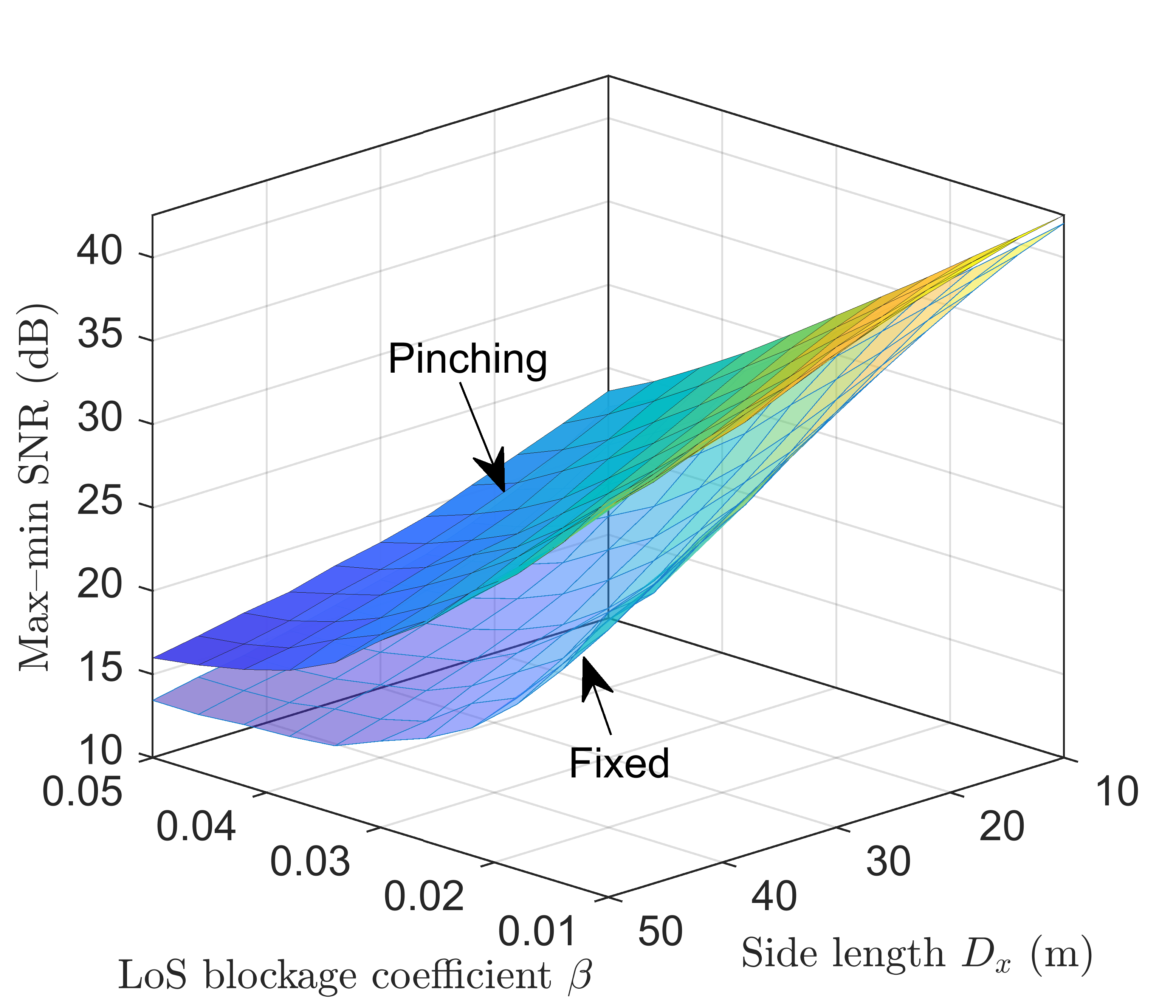}
    \label{fig:snr_3d} 
  }\hfill
  \subfloat[Relative max–min SNR gap between pinching and fixed antennas.]{
    \includegraphics[width=0.82\linewidth]{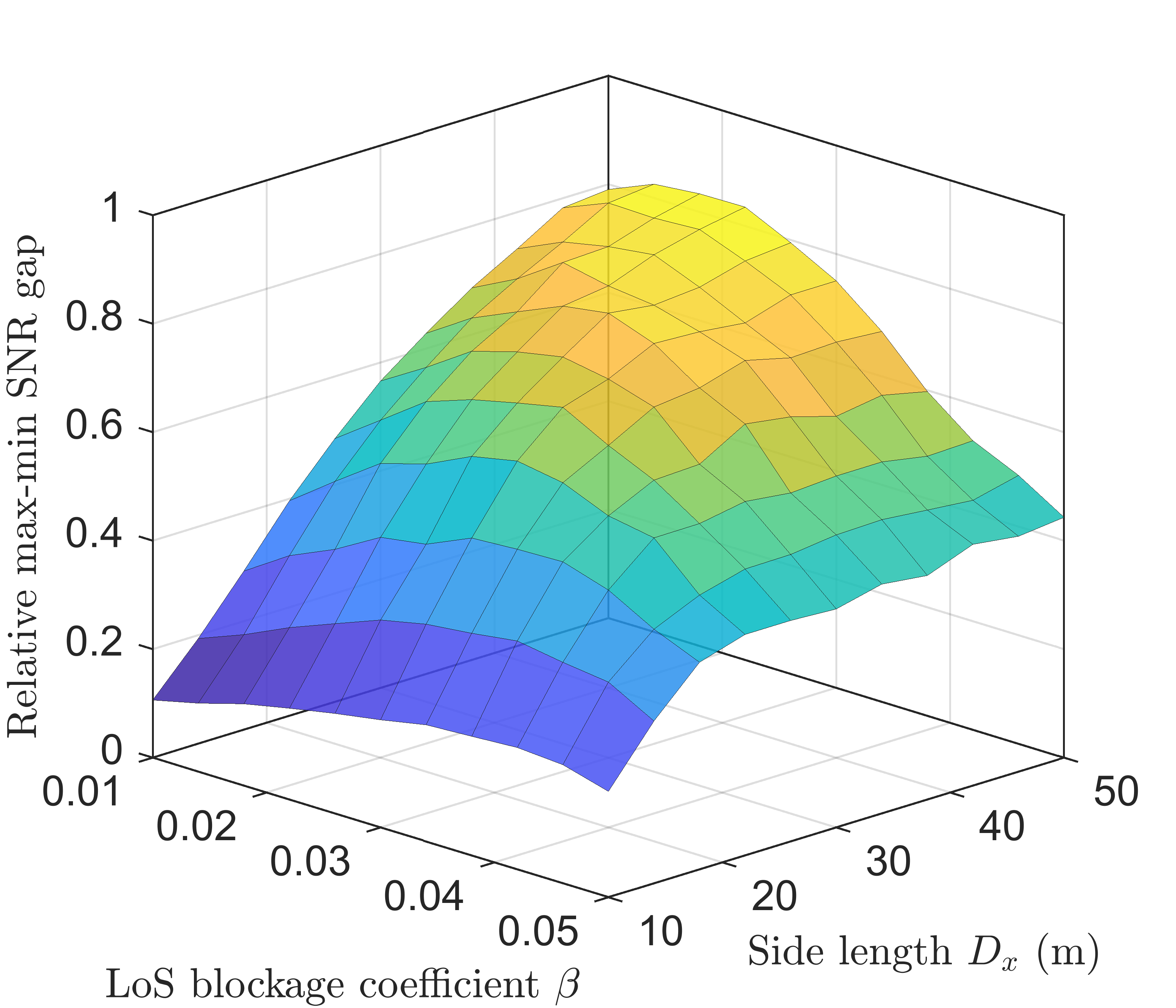} 
    \label{fig:relative_snr_gap_3d}
  }
  \captionsetup{justification=justified, singlelinecheck=false, font=small}	 
  \caption{Max–min average SNR comparison between pinching-antenna and fixed-antenna designs versus  $D_x$ and $\beta$.}  \vspace{-5mm}
  \label{fig:snr_gap_3d}
\end{figure}

Figs. \ref{fig:snr_gap_3d}(a) and \ref{fig:snr_gap_3d}(b) compare the max–min average SNR performance of the proposed pinching-antenna system with that of the fixed-antenna baseline, where the fixed antenna is placed at $[D_x/2, 0, d_v]$. As shown in Fig.~\ref{fig:snr_gap_3d}(a), the pinching-antenna system consistently achieves a higher guaranteed SNR across the whole $(D_x,\beta)$ plane. This gain stems from its ability to adaptively reposition the antenna along the waveguide, thereby shortening the distance to the weakest user and partially compensating for LoS blockage. In contrast, the fixed-antenna system lacks spatial flexibility and thus suffers from more pronounced performance degradation as the communication region expands or the blockage level increases.

\begin{figure}[!t]
  \centering 
  \subfloat[Max–min average SNR of pinching and fixed antennas.]{
    \includegraphics[width=0.82\linewidth]{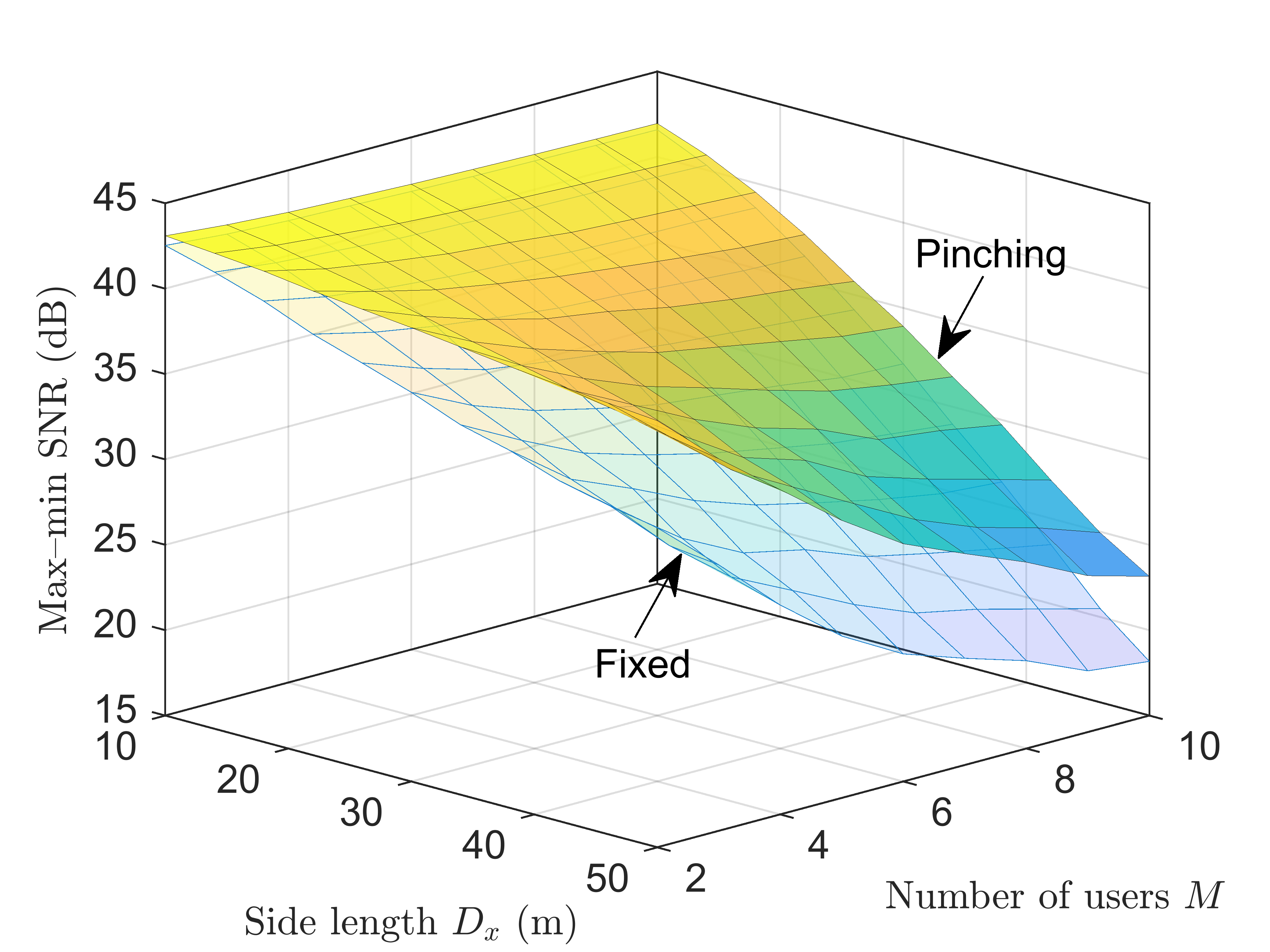}
    \label{fig:snr_M_D} 
  }\hfill \vspace{-3mm}
  \subfloat[Relative max–min average SNR gap between pinching and fixed antennas.]{
    \includegraphics[width=0.82\linewidth]{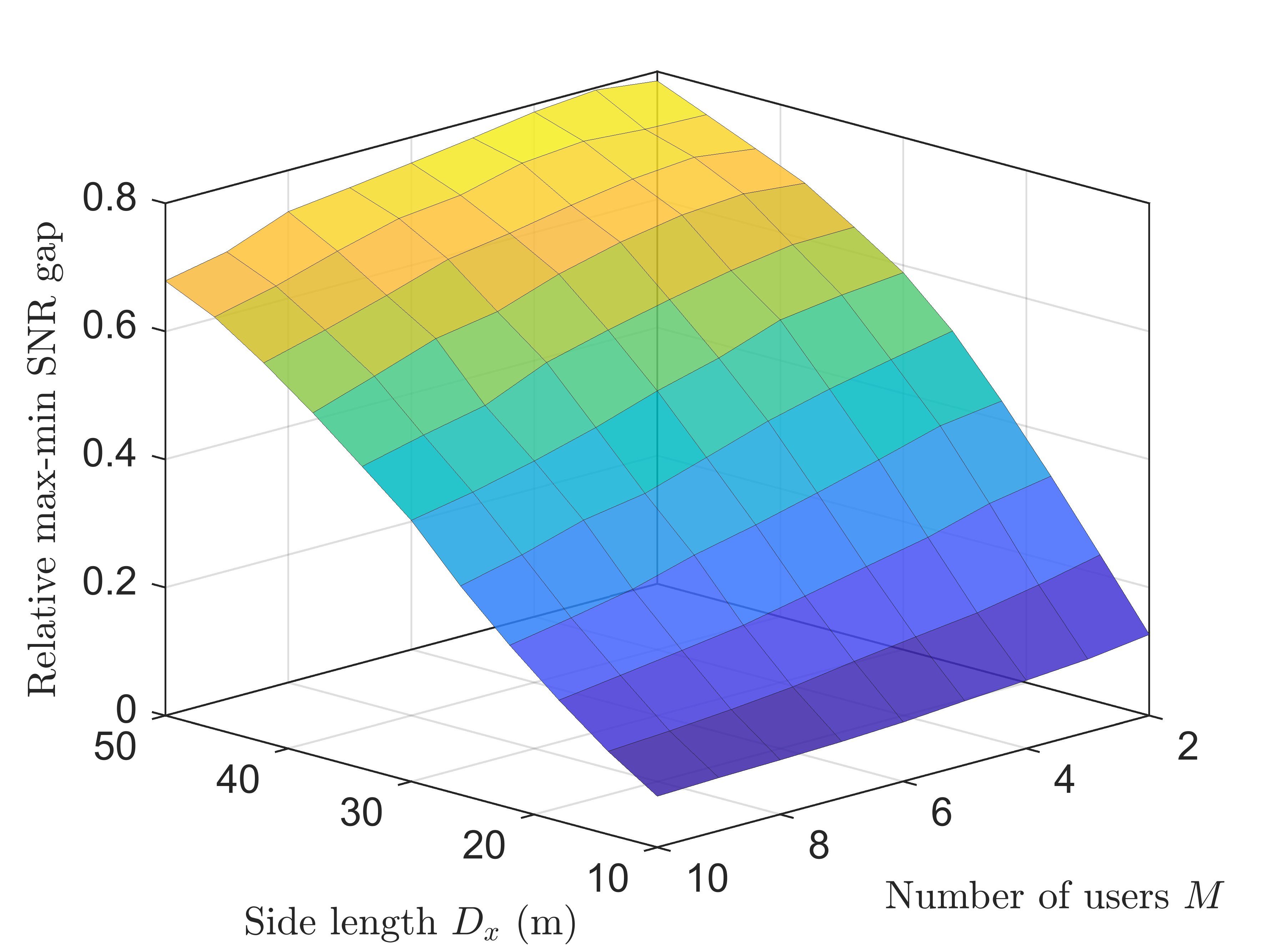} 
    \label{fig:relative_snr_gap_M_D}
  }
  \captionsetup{justification=justified, singlelinecheck=false, font=small}	 
  \caption{Max–min average SNR comparison between pinching-antenna and fixed-antenna designs versus  $D_x$ and $M$.}  \vspace{-5mm}
  \label{fig:snr_gap_M_D}
\end{figure}

Fig.~\ref{fig:snr_gap_3d}(b) illustrates the relative SNR gap, defined as $\frac{\bar \Gamma_{\mathrm{pin}} - \bar \Gamma_{\mathrm{fix}}}{\bar \Gamma_{\mathrm{pin}}}$, where $\bar \Gamma_{\mathrm{pin}}$ and $\bar \Gamma_{\mathrm{fix}}$ denote the max–min average SNR of the pinching-antenna and fixed-antenna schemes, respectively. This metric quantifies the fractional loss incurred when the antenna position is fixed instead of optimally repositioned. The surface reveals that the relative gain is small when either the communication region is very small (limited room to move) or the blockage coefficient is very large (LoS paths are almost always blocked and NLoS dominates). The largest gains occur for large $D_x$ and moderate $\beta$, where user distances are sufficiently diverse and LoS links are still present but vulnerable to blockage, so that adaptive repositioning can substantially improve the weakest links. Overall, these observations highlight that adaptive pinching-antenna placement is particularly beneficial in large and moderately blockage-prone environments, where its spatial flexibility can be fully exploited to enhance both system performance and user fairness.

\begin{figure}[!t]
	\centering
	\includegraphics[width=0.82\linewidth]{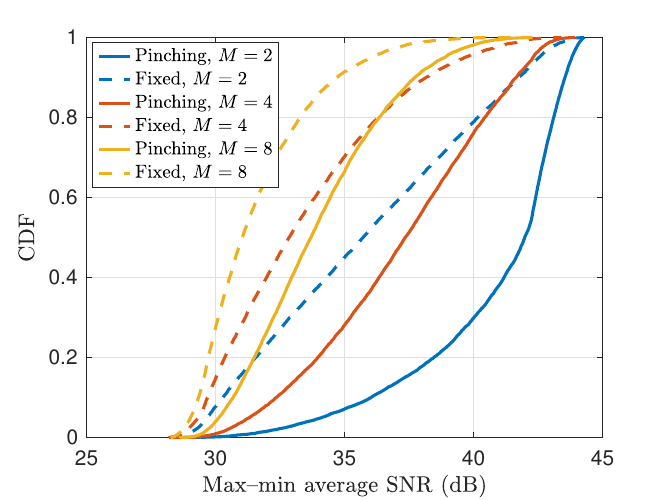}\\
        \captionsetup{justification=justified, singlelinecheck=false, font=small}	
        \caption{CDF of the max-min average SNR for pinching- and fixed-antenna systems with different numbers of users $M$.}  \vspace{-5mm}
    \label{fig:CDF_SNR}
\end{figure}

To gain deeper insights into the impact of user density and region size on system performance, we further evaluate the max--min average SNR of the pinching-antenna system versus the fixed-antenna baseline under varying numbers of users $M$ and side lengths $D_x$ of the communication region, as shown in Fig.~\ref{fig:snr_gap_M_D}. As observed in Fig.~\ref{fig:snr_gap_M_D}(a), the pinching-antenna system consistently outperforms the fixed-antenna design across all tested configurations. Fig.~\ref{fig:snr_gap_M_D}(b) presents the relative max--min average SNR gap as a function of $M$ and $D_x$. The results reveal that the relative gap generally increases with $D_x$, in line with our previous observations that larger regions offer more room to exploit adaptive antenna repositioning. For a given $D_x$, the relative gain initially grows with $M$ but gradually saturates. This behavior arises because, when the number of users is small, the weakest user's location dominates the optimization, allowing repositioning to substantially improve fairness. As $M$ increases and users become more spatially dispersed, the pinching antenna must balance a larger set of links, so the marginal benefit of repositioning for each additional user diminishes, especially in dense deployment scenarios.

The CDFs of the max–min average SNR for the pinching-antenna and fixed-antenna systems are shown in Fig. \ref{fig:CDF_SNR} for different numbers of users $M \in \{2,4,8\}$.
Across all considered values of $M$, the proposed pinching-antenna system consistently achieves a higher max–min average SNR over the whole distribution range compared with its fixed-antenna counterpart, indicating clear gains in both typical and worst-case performance under the random LoS/NLoS channel model.
Moreover, as the number of users increases, the CDF curves of both schemes shift to the left, reflecting the increasing difficulty of balancing the link quality among more spatially distributed users.
Nevertheless, the performance gap between the pinching and fixed antennas remains evident for all $M$, demonstrating that the pinching-antenna gain is robust even in denser multiuser deployments.

\begin{figure}[!t]
	\centering
	\includegraphics[width=0.82\linewidth]{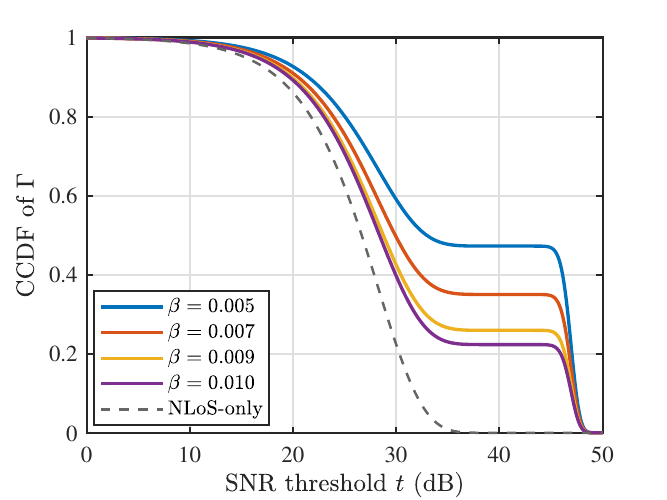}\\
        \captionsetup{justification=justified, singlelinecheck=false, font=small}	
        \caption{CCDF of the instantaneous SNR $\Gamma(t)$ for different LoS blockage coefficients $\beta$ with $D_x = 30$ m.}\vspace{-5mm}
    \label{fig:CCDF}
\end{figure} 
In Fig. \ref{fig:CCDF}, we plot the CCDF of the instantaneous SNR $\Gamma(t)$ under different LoS blockage coefficients $\beta$. In the simulation, we consider a system where one pinching antenna serves one user. The user is fixed at $[10,5,0]$ and the pinching antenna is deployed at $[5,0,10]$. Several observations can be made from Fig. \ref{fig:CCDF}. 
First, for small and moderate thresholds $t$, all curves start close to one and monotonically decrease, as expected. 
Compared with the NLoS-only baseline, the presence of a probabilistic LoS path significantly enhances the SNR tail behaviour, especially in the medium and high SNR regimes. 
Second, for a wide range of thresholds, the CCDF curves with probabilistic LoS exhibit an almost flat plateau, whose level decreases with the blockage coefficient $\beta$. 
This plateau corresponds to the regime where the NLoS component alone cannot support the target SNR, while the deterministic LoS power is still sufficient; in this case, the probability of exceeding the threshold is essentially equal to the LoS probability $p_{\mathrm{LoS}}(r_m)=e^{-\beta r_m^2}$, which directly explains the dependence on $\beta$. 
Finally, when $t$ approaches and exceeds the LoS-limited SNR, i.e., $t \gtrsim \rho_m \eta / r_m^2(\tilde x)$, all curves rapidly drop to zero, since even LoS realizations can no longer satisfy such an overly stringent requirement. 
These behaviours confirm the analytical CCDF in \eqref{eq:CCDF_final_full_new} and illustrate how the blockage coefficient controls the high-SNR reliability of pinching-antenna links under the proposed hybrid LoS/NLoS channel model.

\begin{figure}[!t]
	\centering
	\includegraphics[width=0.82\linewidth]{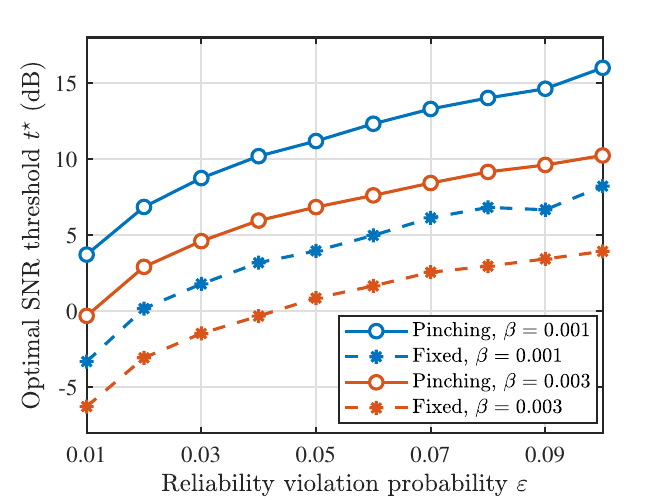}\\
        \captionsetup{justification=justified, singlelinecheck=false, font=small}	
        \caption{Optimal SNR threshold $t^*$ versus outage probability $\varepsilon_m = \varepsilon, \forall m,$ for pinching and fixed antenna systems with $D_x = 40$ m.}\vspace{-3mm} \label{fig:oc t epsilon} 
\end{figure} 

In Fig.~\ref{fig:oc t epsilon}, the optimal SNR thresholds $t^\star$ of the pinching-antenna and fixed-antenna systems are plotted versus the outage probability $\varepsilon$ for two LoS blockage coefficients $\beta$. For both architectures and both blockage levels, $t^\star$ increases monotonically with $\varepsilon$, since relaxing the reliability requirement allows a higher admissible SNR threshold. At a given $\varepsilon$, the achievable $t^\star$ decreases as $\beta$ increases, reflecting the degradation caused by more frequent LoS blockage. Moreover, for any fixed $\beta$ and $\varepsilon$, the pinching-antenna system consistently achieves a higher $t^\star$ than the fixed-antenna benchmark, thanks to its ability to reposition along the waveguide and improve the users' effective channel conditions. Overall, these results provide clear quantitative evidence that pinching antennas can significantly enhance outage-constrained performance and link reliability across a range of blockage regimes.

\begin{figure}[!t]
	\centering
	\includegraphics[width=0.82\linewidth]{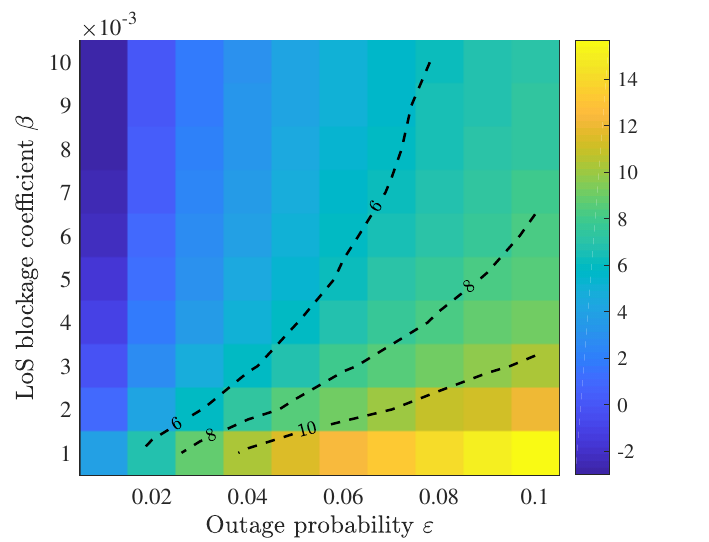}\\
        \captionsetup{justification=justified, singlelinecheck=false, font=small}	
        \caption{Optimal outage-constrained SNR threshold $t^\star$ of the pinching-antenna system versus $\beta$ and  $\varepsilon_m = \varepsilon, \forall m$, with contour lines marking representative SNR levels.} \vspace{-5mm} \label{fig:heatmap_outage} 
\end{figure} 

Fig.~\ref{fig:heatmap_outage} presents a heatmap with contour lines illustrating the optimal outage-constrained SNR threshold $t^\star$ of the pinching-antenna system as a function of the LoS blockage coefficient $\beta \in [10^{-3},10^{-2}]$ and the target outage probability $\varepsilon \in [0.01,0.1]$. The color shading indicates the achievable SNR threshold, while the dashed contour curves mark representative $t^\star$ levels for visual clarity.

As observed, a larger tolerable outage probability $\varepsilon$ allows the system to support a higher SNR threshold $t^\star$, since the design no longer needs to guarantee extremely high reliability for every channel realization. In contrast, $t^\star$ decreases with increasing blockage coefficient $\beta$, reflecting the degradation of the LoS component under more frequent or severe obstructions. The contour pattern further reveals an inherent tradeoff between reliability and propagation conditions: for higher $\beta$, achieving the same SNR level requires relaxing the reliability target, i.e., operating at a larger outage probability $\varepsilon$. This illustrates how environmental blockage and reliability requirements jointly shape the achievable performance of pinching-antenna systems.

\begin{figure}[!t]
	\centering
	\includegraphics[width=0.82\linewidth]{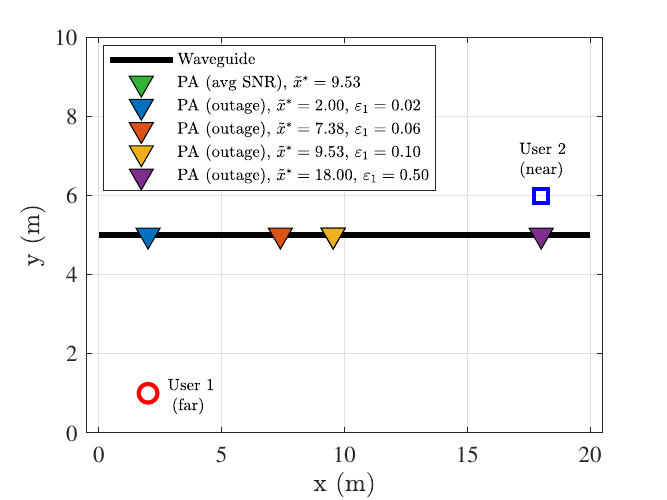}\\
        \captionsetup{justification=justified, singlelinecheck=false, font=small}	
        \caption{Top view of the communication area showing the optimized pinching-antenna positions under the average-SNR and outage-constrained designs.} \vspace{-5mm}
    \label{fig:top view}
\end{figure} 
Fig. \ref{fig:top view} illustrates the optimized pinching-antenna positions obtained under the average-SNR-based and outage-constrained designs in a two-user scenario, where one user (User 1) is positioned far from the waveguide and the other (User 2) is located close to it. For the outage-constrained case, the reliability requirement of the near user is fixed at $\varepsilon_2 = 0.1$, whereas that of the far user $\varepsilon_1$ takes values in $\{0.02, 0.06, 0.10, 0.50\}$. The corresponding pinching-antenna positions are plotted to illustrate how the optimal antenna location adapts to different reliability constraints.
As observed, under the average-SNR-based design, the pinching antenna is positioned near the midpoint between the two users to balance their link qualities and maximize the minimum average SNR. In contrast, the outage-constrained design produces antenna placements that shift toward the user with the tighter reliability requirement. When $\varepsilon_1$ is very small (e.g., $\varepsilon_1=0.02$), the pinching antenna moves close to the far user to reduce its outage probability. As $\varepsilon_1$ increases and eventually becomes much larger than $\varepsilon_2$ (e.g., $\varepsilon_1=0.50$), the antenna moves toward the near user. This behavior highlights the distinct design philosophies of the two metrics: the average-SNR-based optimization promotes long-term fairness, whereas the outage-constrained scheme prioritizes reliability fairness across users under probabilistic channel variations.

\vspace{-3mm}

\section{Conclusion} \label{sec:conclusions}
This paper investigated the performance optimization of multiuser pinching-antenna systems under a realistic probabilistic channel model incorporating both random LoS blockage and NLoS scattering effects. Two complementary design metrics were introduced to jointly capture long-term efficiency and short-term reliability aspects: the average-SNR-based metric and the outage-constrained metric. Based on these metrics, we formulated two optimization problems: one maximizing the max-min average SNR across users, and the other maximizing a guaranteed SNR threshold under per-user outage constraints.
Although both problems are inherently nonconvex, we revealed their hidden monotonic structures and proposed efficient bisection-based algorithms that globally solve them with very low computational complexity. The developed methods require only scalar evaluations and avoid any high-dimensional search or matrix operations, making them highly scalable and suitable for real-time implementation. Simulation results validated the optimality and efficiency of the proposed algorithms, and further demonstrated that pinching-antenna systems can substantially outperform conventional fixed-antenna schemes even in the presence of LoS blockage and NLoS fading. The results of this work not only highlight the potential of pinching antennas for robust and adaptive wireless communication but also provide a unified framework for reliability-aware and efficiency-oriented system design. Future work may extend these formulations to non-orthogonal multiple access, multi-waveguide architectures, and integrated sensing and communication applications.

\vspace{-3mm}

\begin{appendices}
\section{Proof of Proposition \ref{prop:CCDF_new}}\label{app: ccdf_new}

First, recall the channel model with probabilistic LoS and random NLoS components:
\begin{align}
    h_m(\tilde x) = \gamma_m h_m^{\mathrm{LoS}}(\tilde x) + h_m^{\mathrm{NLoS}}(\tilde x).
\end{align}
The instantaneous SNR is $\Gamma_m(\tilde x)=\rho_m |h_m(\tilde x)|^2$.
Using the law of total probability over $\gamma_m$, we obtain
\begin{small}
\begin{align}
&\Pr\big(\Gamma_m(\tilde x)\ge t\big)\notag\\
&= \Pr(\gamma_m=1)\,\Pr\!\Big(\rho_m\big|h_m^{\mathrm{LoS}}(\tilde x)+h_m^{\mathrm{NLoS}}(\tilde x)\big|^2\ge t\;\Big|\;\gamma_m=1\Big)\notag\\
&\quad+\Pr(\gamma_m=0)\,\Pr\!\Big(\rho_m\big|h_m^{\mathrm{NLoS}}(\tilde x)\big|^2\ge t\;\Big|\;\gamma_m=0\Big)\notag\\
&= e^{-\beta r_m^2(\tilde x)} \underbrace{\Pr \Big(\big|h_m^{\mathrm{LoS}}(\tilde x)+h_m^{\mathrm{NLoS}}(\tilde x)\big|^2\ge \tfrac{t}{\rho_m}\Big)}_{\mathbb{P}_1}\notag\\
&\quad +\big(1-e^{-\beta r_m^2(\tilde x)}\big)  \underbrace{\Pr \Big(\big|h_m^{\mathrm{NLoS}}(\tilde x)\big|^2\ge \tfrac{t}{\rho_m}\Big)}_{\mathbb{P}_2}.
\label{eq:ccdf-mixture-new}
\end{align}
\end{small}%

The probability $\mathbb{P}_1$ corresponds to the event that the instantaneous received power under LoS conditions exceeds a threshold. 
To characterize this term, we factor out the distance-dependent attenuation:
\begin{align}
    h_m^{\mathrm{LoS}}(\tilde x) + h_m^{\mathrm{NLoS}}(\tilde x)
    = \frac{1}{r_m(\tilde x)}\Big(\sqrt{\eta} e^{-j\phi_m(\tilde x)} + \mu_m Z\Big),
\end{align}
where $Z \sim \mathcal{CN}(0,1)$ is a standard circularly symmetric complex Gaussian variable.
Define
\begin{align}
    H' \triangleq \sqrt{\eta} e^{-j\phi_m(\tilde x)} + \mu_m Z,
    \qquad R' \triangleq |H'|.
\end{align}
Then $H'$ is a Rician random variable with noncentrality $\sqrt{\eta}$ and per-component variance $\sigma^2=\mu_m^2/2$, while the actual channel magnitude under LoS is
\begin{align}
    R \triangleq |h_m^{\mathrm{LoS}}(\tilde x) + h_m^{\mathrm{NLoS}}(\tilde x)| = \frac{R'}{r_m(\tilde x)}.
\end{align}
Since the instantaneous channel power satisfies $|h|^2 = R^2$, the event 
$\{|h|^2 \ge t/\rho_m\}$ is equivalent to $\{R \ge \sqrt{t/\rho_m}\}$, or equivalently
\begin{align}
    R' \ge \sqrt{\tfrac{t}{\rho_m}}\,r_m(\tilde x).
\end{align}

To express the resulting probability in a standard form, we normalize the variable $R'$ by the standard deviation $\sigma$ of the NLoS component. 
Define $z = R'/\sigma$, so that $R' = \sigma z$ and $dR' = \sigma dz$. 
Using the Rician PDF of $R'$ \cite{simon2004digital}, we obtain the PDF of $z$ as
\begin{align}
    f_Z(z) = z \exp\!\Big(-\frac{z^2+a_m^2}{2}\Big) I_0(a_m z),\quad z\ge 0,
\end{align}
where
\begin{align}
    a_m = \frac{\sqrt{\eta}}{\sigma}
        = \sqrt{2}\,\frac{\sqrt{\eta}}{\mu_m}.
\end{align}
On the other hand, the threshold on $R'$ translates into a threshold on $z$:
\begin{align}
    z \ge b_m(r_m(\tilde x),t)
    &\triangleq \frac{\sqrt{t/\rho_m}\,r_m(\tilde x)}{\sigma} \notag\\
    &= \sqrt{2}\,\frac{r_m(\tilde x)\sqrt{t/\rho_m}}{\mu_m}.
\end{align}
Therefore, by using the definition of the Marcum-$Q$ function \cite{kapinas2009monotonicity}, we have
\begin{subequations} \label{eqn: auxilliary a b new}
    \begin{align}
        a_m &= \sqrt{2}\,\frac{\sqrt{\eta}}{\mu_m},\\
        b_m(r_m(\tilde x),t) &= \sqrt{2}\,\frac{r_m(\tilde x)\sqrt{t/\rho_m}}{\mu_m},
    \end{align}
\end{subequations}
and
\begin{align}
    &\Pr \Big(\big|h_m^{\mathrm{LoS}}(\tilde x) + h_m^{\mathrm{NLoS}}(\tilde x)\big|^2 \ge \tfrac{t}{\rho_m}\Big) \notag\\
    &=\int_{b_m(r_m(\tilde x),t)}^\infty z\,\exp\!\Big(-\frac{z^2+a_m^2}{2}\Big)\,I_0\!\big(a_m z\big)\,dz \notag \\
    &= Q_1\!\big(a_m,\,b_m(r_m(\tilde x),t)\big). \label{eq:rician-mq-new}
\end{align} 

The second term, $\mathbb{P}_2$, corresponds to the pure NLoS case, where 
\begin{align}
    h_m^{\mathrm{NLoS}}(\tilde x)\sim\mathcal{CN}\!\Big(0,\frac{\mu_m^2}{r_m^2(\tilde x)}\Big).
\end{align}
In this case, the real and imaginary parts are distributed as $\mathcal{N}\big(0,\mu_m^2/(2r_m^2(\tilde x))\big)$, and thus $|h_m^{\mathrm{NLoS}}(\tilde x)|^2$ follows an exponential distribution with mean $\mu_m^2 / r_m^2(\tilde x)$. This yields
\begin{equation}
    \Pr\!\Big(\big|h_m^{\mathrm{NLoS}}(\tilde x)\big|^2 \ge \tfrac{t}{\rho_m}\Big)
    = \exp\!\Big(-\frac{t\,r_m^2(\tilde x)}{\rho_m\mu_m^2}\Big).
    \label{eq:nlos-exptail-new}
\end{equation}
Finally, substituting \eqref{eq:rician-mq-new} and \eqref{eq:nlos-exptail-new} into \eqref{eq:ccdf-mixture-new} yields the closed-form CCDF expression in \eqref{eq:CCDF_final_full_new}.

The remaining task is to show that, for any fixed $t$, $\Pr(\Gamma_m(\tilde x) \ge t)$ is a strictly decreasing function of the user-to-antenna distance. 
To proceed, we first recall the monotonic properties of the first-order Marcum-$Q$ function $Q_1(a,b)$: for any $a,b \ge 0$, $Q_1(a,b)$ is strictly increasing in $a$ for any fixed $b$ and strictly decreasing in $b$ for any fixed $a$ \cite{kapinas2009monotonicity}. 
Meanwhile, from the definitions in \eqref{eqn: auxilliary a b new}, we observe that $a_m$ is a positive constant, while $b_m(r_m(\tilde x),t)$ is strictly increasing in the distance $r_m(\tilde x)$ (and hence in $r_m^2(\tilde x)$) for any fixed $t>0$. 
Therefore, for fixed $t$, the composition $Q_1\big(a_m,b_m(r_m(\tilde x),t)\big)$ is strictly decreasing in $r_m(\tilde x)$.

Let $y \triangleq r_m^2(\tilde x)$ for brevity. Using \eqref{eq:CCDF_final_full_new}, we can rewrite the CCDF as \vspace{-3mm}

\begin{small}
\begin{align}
    &\Pr(\Gamma_m(\tilde x)\ge t) \notag\\
    & = e^{-\beta y}\, Q_1\!\big(a_m,b_m(\sqrt{y},t)\big) 
        \!+\! \big(1\!-\!e^{-\beta y}\big)\exp\!\Big(\!-\frac{t\,y}{\rho_m\mu_m^2}\Big).
\end{align}
\end{small}%
On the other hand, by using the monotonicity of the Marcum-$Q$ function, we have
\begin{align}
    Q_1\big(a_m,b_m(\sqrt{y},t)\big) 
    &\geq Q_1\big(0,b_m(\sqrt{y},t)\big) \notag\\
    &= \exp\!\Big(-\frac{b_m^2(\sqrt{y},t)}{2}\Big) \notag\\
    &= \exp\!\Big(-\frac{t\,y}{\rho_m\mu_m^2}\Big).
\end{align}
Therefore, the difference term $Q_1\big(a_m,b_m(\sqrt{y},t)\big) - \exp\big(-\frac{t\,y}{\rho_m\mu_m^2}\big)$ is strictly positive. We can thus rewrite the CCDF as
\begin{align}
    &\Pr(\Gamma_m(\tilde x)\ge t) \notag\\
    &= e^{-\beta y}\!\Big[ Q_1\!\big(a_m,b_m(\sqrt{y},t)\big) 
    - \exp\!\Big(-\frac{t\,y}{\rho_m\mu_m^2}\Big) \Big] \notag\\
    &\quad + \exp\!\Big(-\frac{t\,y}{\rho_m\mu_m^2}\Big).
\end{align}

Here, both $e^{-\beta y}$ and $\exp\!\big(-\frac{t\,y}{\rho_m\mu_m^2}\big)$ are strictly decreasing in $y$, whereas the second exponential term is independent of $a_m$ and $b_m(\cdot)$ and serves as a baseline.  
At the same time, the bracketed term is strictly positive and strictly decreasing in $y$, since $Q_1\!\big(a_m,b_m(\sqrt{y},t)\big)$ is strictly decreasing in $y$ and dominates the exponential term from below.
Consequently, the product between two strictly positive and strictly decreasing functions, i.e., $e^{-\beta y}$ and $\big[ Q_1(a_m,b_m(\sqrt{y},t)) - \exp(-t y/(\rho_m\mu_m^2)) \big]$, is also strictly decreasing in $y$.  
Adding the strictly decreasing term $\exp(-t y/(\rho_m\mu_m^2))$ preserves strict monotonicity, and therefore $\Pr(\Gamma_m(\tilde x)\ge t)$ strictly decreases with $y=r_m^2(\tilde x)$ for any fixed threshold $t>0$. Since $r_m(\tilde x)$ is strictly increasing in $y$, this implies that the CCDF is strictly decreasing with the user-to-antenna distance.  
This completes the proof of Proposition \ref{prop:CCDF_new}.
\hfill$\blacksquare$

\end{appendices}


\smaller[1]

\end{document}